\documentclass[twocolumn,showpacs,preprintnumbers,amsmath,amssymb,floatfix]{revtex4-1}

\usepackage{graphicx}
\usepackage{dcolumn}
\usepackage{bm}
\usepackage{color}
\usepackage{url}
\usepackage{amsmath}

\graphicspath{{./figures/}}
\newcommand{\be}{\begin{equation}}
\newcommand{\ee}{\end{equation}}
\newcommand{\ba}{\begin{eqnarray}}
\newcommand{\ea}{\end{eqnarray}}
\newcommand{\Mc}{{\cal M}}
\newcommand{\Ms}{M_{\odot}}

\def\ltsima{$\; \buildrel < \over \sim \;$}
\def\simlt{\lower.5ex\hbox{\ltsima}}
\def\gtsima{$\; \buildrel > \over \sim \;$}
\def\simgt{\lower.5ex\hbox{\gtsima}}
\begin{document}

\title[Testing General Relativity using Bayesian model selection]{Testing General Relativity using Bayesian model selection: Applications to observations of gravitational waves from compact binary systems}

\author{Walter Del Pozzo$^{1,2}$, John Veitch$^{1,3}$ and Alberto Vecchio$^1$}
\affiliation{$^1$School of Physics and Astronomy, University of Birmingham, 
  Edgbaston, Birmingham B15 2TT, UK\\ $^2$Nikhef � National Institute for Subatomic Physics, Science Park 105, 1098 XG Amsterdam, The Netherlands \\$^3$Physics and Astronomy, Cardiff University, Queen's Buildings, The Parade, Cardiff CF24 3AA, UK}
\date{\today}

\begin{abstract}
Second generation interferometric gravitational wave detectors, such as Advanced LIGO and Advanced Virgo, are expected to begin operation by 2015. Such instruments plan to reach sensitivities that will offer the unique possibility to test General Relativity in the dynamical, strong field regime and investigate departures from its predictions, in particular using the signal from coalescing binary systems. We introduce a statistical framework based on Bayesian model selection in which the Bayes factor between two competing hypotheses measures which theory is favored by the data. Probability density functions of the model parameters are then used to quantify the inference on individual parameters. We also develop a method to combine the information coming from multiple independent observations of gravitational waves, and show how much stronger inference could be.
As an introduction and illustration of this framework -- and a practical numerical implementation through the Monte Carlo integration technique of nested sampling -- we apply it to gravitational waves from the inspiral phase of coalescing binary systems as predicted by General Relativity and a very simple alternative theory in which the graviton has a non-zero mass. This method can trivially (and should) be extended to more realistic and physically motivated theories.
\end{abstract}


\maketitle

\section{Introduction}

General Relativity (GR) has so far passed every experimental test with flying colours. Yet, a whole range of efforts is under way to put Einstein's theory under even more intense experimental scrutiny in the coming years~\cite{Will:2006}. Amongst these, highly sensitive gravitational wave experiments are opening new means to probe gravity in the dynamical, strong-field regime~\cite{SathyaprakashSchutz:2009}. Ground-based gravitational-wave laser interferometers, LIGO~\cite{BarishWeiss:1999,lsc-s5-instr}  and Virgo~\cite{virgo} have now reached a sensitivity that could plausibly lead to the first direct detection of gravitational waves. The upgrade of these instruments to the second generation (also known as advanced configuration) is already under way; Advanced LIGO~\cite{advligo} and Advanced Virgo~\cite{advvirgo} are expected to start science observations by 2015, and to provide a wealth of detections from a variety of sources~\cite{CutlerThorne:2002,Kokkotas:2008,lvc-cbc-rates}. As soon as a positive detection is achieved, one can surely expect that gravitational-wave data will be used to test the predictions of General Relativity, see~\cite{Will:2006,SathyaprakashSchutz:2009} for recent reviews. In the longer term future, very-high sensitivity laser interferometers, both space-based -- such as the Laser Interferometer Space Antenna (LISA)~\cite{lisa} and Decigo~\cite{decigo} -- and ground-based third-generation instruments, such as the Einstein gravitational-wave Telescope (ET)~\cite{FreiseEtAl:2009, HildEtAl:2009,PunturoEtAl:2010}, will increase our ability to test alternative theories of gravity.

Tests of General Relativity through gravitational wave observations have already been discussed in several studies~\cite{Will:1994,Ryan:1997,Will:1998,Will:2003,WillYunes:2004,BertiBuonannoWill:2005,GlampedakisBabak:2005,KesdenGairKamionkowski:2005,Hughes:2006,ArunEtAl:2006,BertiCardosoWill:2006,BarackCutler:2007,AlexanderFinnYunes:2008,GairLiMandel:2008,YunesFinn:2009,ArunWill:2009,StavridisWill:2009,YunesSopuerta:2009,YunesPretorius:2009,ApostolatosEtAl:2009,Schutz:2009,KeppelAjith:2010}. Of particular interest are observations of the coalescence of binary systems, as they allow us to probe the dynamic, highly relativistic and strong-field regime through the precise monitoring of the amplitude and phase evolution of gravitational waves, that can be accurately predicted in General Relativity and alternative theories of gravity.

The conceptual approach that most of these studies have employed to investigate the ability to test General Relativity is based on the computation of the expected accuracy of the measurements of the unknown parameters that characterise the radiation, including those introduced by alternative theories of gravity; moreover, the statistical errors are estimated using the inverse of the Fisher information matrix~\cite{JaynesBretthorst:2003}. These results are useful as approximate figures of merit for the constraints that can be expected to be derived from observations. However, they suffer from various conceptual (and practical) limitations that we address in this paper. The first is that at the conceptual level, those studies do not actually address whether observations will be able to discriminate an alternative theory of gravity from General Relativity. They simply \emph{assume} that the true theory of gravity is different from General Relativity, and by computing the expected statistical errors on the unknown signal parameters (including those that encode deviations from GR) make some statements on whether or not the observations have sensitivity to the relevant parameter(s). [This leaves aside the issue that the computation of the variance-covariance matrix simply provides a lower bound, the Cramer-Rao bound~\cite{Rao:1945,Cramer:1946}, to the variance of the statistical errors, which is a meaningful bound on the accuracy of parameter estimation only in the limit of high signal-to-noise ratio, see \emph{e.g.} Refs.~\cite{NicholsonVecchio:1998,BalasubramanianDhurandhar:1998,Vallisneri:2008,ZanolinVitaleMakris:2009}. Secondly, those studies ignore what would be the consequences of (small) deviations from GR, such that detections can still be achieved using GR waveforms when the actual theory of gravity differs from GR, but parameter estimation can be affected due to the use of the ``wrong" waveform model; this issue has been recently discussed in Ref.~\cite{YunesPretorius:2009} and termed ``bias in gravitational wave astronomy". In addition to this, past studies do not take into account the fact that one can take advantage of the (hopefully) many detections to provide better constraints on alternative theories of gravity by combining the observations. In fact, although each astrophysical source will be characterised by different astrophysical parameters (such as masses and spins, in the case of coalescing binaries), if there is a deviation from GR described by some ``global" fundamental parameters of the new theory that are the same for every system -- \emph{e.g.} the gravitational radiation dispersion velocity is affected in such a way that one can associate the graviton a mass different from zero, corresponding to a specific value -- then one can \emph{combine}, see \emph{e.g.} Ref.~\cite{Mandell:2010} all the observations and obtain better constraints. Finally, there has been no actual attempt to provide a method to address these issues in an analysis that can be implemented in practice and therefore move towards using actual gravitational wave data to place constraints on theories of gravity.

In this paper we tackle these specific issues: we introduce a conceptual and practical approach within the framework of Bayesian inference to discriminate between different theories of gravity by performing model selection, and to estimate the unknown model parameters. The approach based on Bayesian inference is particularly simple and powerful as it provides both a statement on the relative probabilities of models (a given alternative theory of gravity versus General Relativity) and on the distribution of the unknown parameters that characterise the theory, the (marginalised) posterior probability density functions (PDFs). The conceptual simplicity of Bayes' theorem is balanced by the
computationally expensive $N$-dimensional integral (where $N$, usually $\simgt 10$, is the total number of the unknown parameters) on which this theorem relies for the calculation of the evidence and the marginalised PDF's. Several integration techniques, such (Reversible Jump) Markov-chain Monte Carlo methods~\cite{mcmc-in-practice,Green:1995}, thermodynamic integration~\cite{GelmanMeng:1998} and Nested Sampling~\cite{Skilling:AIP} have been explored in a range of fields to tackle this computational challenge. For our analysis we use a nested sampling algorithm that some of us have developed for applications in the context of observations of coalescing binaries with ground-based instruments~\cite{VeitchVecchio:2010}, and that has been shown to allow such $N$-dimensional integrals to be computed in an efficient and relatively computationally inexpensive way~\cite{VeitchVecchio:2010,VeitchVecchio:2008a, VeitchVecchio:2008b,AylottEtAl:2009,AVV:2009}.

The method that we present here is completely general and can be applied to observations of any gravitational wave source and any alternative theory of gravity. However, for the purpose of introducing and illustrating the method, in this paper we focus specifically on the case of a ``massive graviton'' theory --\emph{i.e.} a theory of gravity in which the boson mediating the gravitational interaction is characterised by a rest-mass $m_\mathrm{g}$ different from zero, and the corresponding Compton wavelength of the graviton $\lambda_\mathrm{g}$ is finite (the GR case corresponds to $m_\mathrm{g} = 0$ and $\lambda_\mathrm{g} \rightarrow \infty$) -- and consider how one would go about testing GR against this alternative theory in a statistically rigorous way, by providing both a conceptual and practical (in the sense that is readily applicable to gravitational wave observations) approach to this problem. The reason for choosing a massive graviton theory for our proof-of-concept analysis is two-fold: (i) the gravitational radiation emitted by coalescing compact binaries in a massive graviton theory takes a particularly simple form characterised by only one additional unknown parameter -- the Compton wavelength of the graviton $\lambda_\mathrm{g}$ -- and therefore provides an ideal proof-of-concept case to study, and (ii) several studies have already explored the feasibility of placing new limits on the graviton mass using gravitational wave observations, and it is therefore useful to compare those expectations with actual results from a rigorous statistical analysis performed on mock data sets. Clearly the analysis that we show here can be applied to any other theory of gravity.

The paper is organised as follows. In Section \ref{s:method} we discuss the statistical method that we employ. In Section \ref{s:model} we present the models we use as a test case for our study, and introduce the gravitational waveform generated by in-spiralling compact binaries in General Relativity, and in a``massive graviton'' theory. Sections~\ref{s:results-single} and~\ref{s:results-combined} contain the main results of the paper: in Section \ref{s:results-single} we show the results of our analysis for the observation of a single gravitational wave signal; in Section \ref{s:results-combined} we derive a method for combining multiple observations which are expected from advanced interferometers, and we show how effective it can be for the specific example at hand to further constrain $\lambda_\mathrm{g}$. Finally in Section \ref{s:conclusions} we summarise our work.

\section{Method}
\label{s:method}

Let us consider a set of theories of gravity $\{{\cal H}_j\}$, including General Relativity that we want to test using observations of gravitational waves emitted during the coalescence of binary systems of compact objects, either black holes or neutron stars. Each theory makes a prediction on the gravitational waveform $h(t; \vec{\theta})$, that depends on the specific theory ${\cal H}_j$ and a set of unknown parameters $\vec{\theta}$. The statements that we can make on any given theory is based on a data set $d$ (observations) and all the relevant prior information $I$ that we hold. 

Within the framework of Bayesian inference, the key quantity that one needs to compute is the posterior probability of a given theory (a ``model'' or hypothesis) ${\cal H}_j$. Applying Bayes' theorem we obtain
\be
P({\cal H}_j | d, I) = \frac{P({\cal H}_j | I)\,P(d |{\cal H}_j, I)}{P(d | I)}\,,
\label{posterior}
\ee
where $P({\cal H}_j | d, I)$ is the \emph{posterior probability} of the model ${\cal H}_j$ given the data, $P({\cal H}_j | I)$ is the \emph{prior probability} of hypothesis ${\cal H}_j$, and $P(d | {\cal H}_j, I)$ is the \emph{marginal likelihood} or \emph{evidence} for ${\cal H}_j$ that can be written as:
\ba
P(d | {\cal H}_j, I) & = & {\cal L}({\cal H}_j) 
\nonumber\\
& = & \int d\vec{\theta}\, p(\vec{\theta} | {\cal H}_j, I)\, p(d | \vec{\theta}, {\cal H}_j, I)\,.
\label{e:marglikelihood}
\ea
In the previous expression $p(\vec{\theta} | {\cal H}_j, I)$ is the prior probability density of the unknown parameter vector $\vec{\theta}$ within the theory ${\cal H}_j$ and $p(d | \vec{\theta}, {\cal H}_j, I)$ is the likelihood function of the observation $d$, assuming a given value of the parameters $\vec{\theta}$ and the theory ${\cal H}_j$.  

If we want to compare different models -- for this paper, we concentrate on General Relativity versus an alternative theory of gravity -- in light of the observations made, we can compute the relative posterior probabilities, which is known as the \emph{odds ratio}
\ba\label{e:bayes}
O_{i,j} & = & \frac{P({\cal H}_i|d)}{P({\cal H}_j|d)} 
\nonumber\\
& = & \frac{P({\cal H}_j)}{P({\cal H}_i)}\frac{P(d | {\cal H}_i)}{P(d | {\cal H}_j)}
\nonumber\\
& = &  \frac{P({\cal H}_j)}{P({\cal H}_i)}\,B_{i,j}\,,
\ea
where $P({\cal H}_j)/P({\cal H}_i)$ is the \emph{prior odds} of the two hypotheses, the confidence we assign to the models before any observation, and $B_{i,j}$ is the \emph{Bayes factor}. Here, we are interested in what the data can tell us about the relative probabilities of two models, and so we will not involve the prior odds any further.

In addition to computing the relative probabilities of different theories, one usually wants to make inference on the unknown parameters, and therefore one needs to compute the joint posterior probability density function 
\be
p(\vec{\theta} | d, {\cal H}_j, I) = \frac{p(\vec{\theta} | {\cal H}_j, I) p(d | \vec{\theta}, {\cal H}_j, I)}{p(d|{\cal H}_j, I)}\,.
\label{e:pdf0}
\ee
From the previous expression it is simple to compute the marginalised PDF on any given parameter, say $\theta_1$ within a given theory of gravity ${\cal H}_j$ 
\be
p(\theta_1 | d, {\cal H}_j, I) = \int d\theta_2 \dots \int d\theta_N p(\vec{\theta} | d, {\cal H}_j, I)\,.
\label{e:pdf}
\ee
The key quantities for Bayesian inference in Eq.~(\ref{e:bayes}),~(\ref{e:pdf0}) and~(\ref{e:pdf}) can be efficiently computed using \emph{e.g.} a nested sampling algorithm~\cite{Skilling:AIP}. In this paper use a specific implementation of this technique that we have developed for ground-based observations of coalescing binaries, whose technical details are described in Ref.~\cite{VeitchVecchio:2010}.

\section{Models}
\label{s:model}

In this Section we introduce and review the gravitational waveform approximations that we use, and spell out the models that we consider in the proof-of-concept analyses, whose results are presented in Sections~\ref{s:results-single} and~\ref{s:results-combined}.

\subsection{Gravitational waveforms}

In General Relativity, gravitational waves generated during the in-spiral of compact binary systems are accurately modelled using the post-Newtonian approach, see \emph{e.g.} Ref.~\cite{Blanchet:2006} for a review. Here we use the standard restricted post-Newtonian approximation to the radiation, computing directly the waveform in the frequency domain by taking advantage of the stationary phase approximation. The amplitude of the radiation contains therefore the leading order Newtonian contribution and higher order post-Newtonian terms are retained only in the phase. We further consider the expansion to post$^2$-Newtonian order and we assume that the compact objects have no spins. In summary, we will consider the frequency domain GW signal in the General Relativity case to be
\be
h_\mathrm{GR}(f)={\cal A}f^{-7/6}e^{i\Psi_\mathrm{GR}(f)}\,,
\label{e:hGR}
\ee
where
\be
{\cal A}=\frac{1}{\sqrt{30}\pi^{2/3}}\frac{\Mc^{5/6}}{D_L}\,,
\label{A:gw}
\ee
is the amplitude of the signal from a source at luminosity distance $D_L$, and the phase $\Psi_\mathrm{GR}(f)$ at the post$^2$-Newtonian order is given by~\cite{BlanchetEtAl:1995}:
\ba
\Psi_\mathrm{GR}(f)  & = 2\pi f t_c-\Phi_c+\frac{3}{128}(\pi M f)^{-5/3}\eta^{-1}\Big[1 + \nonumber &\\
 & + \left(\frac{3715}{756}+\frac{55}{9}\eta \right)(\pi M f)^{2/3}-16\pi(\pi M f)+\nonumber & \\
 &  +\left(\frac{15293365}{508032}+\frac{27145}{504}\eta+\frac{3085}{72}\eta^2 \right)(\pi M f)^{4/3}\Big]\,.&
\label{e:phaseGR}
\ea
In the previous expressions $f$ is the GW frequency, 
\ba
M & = & m_1+m_2 \\
\eta & = & \frac{m_1 m_2}{M^2} \\
\Mc & = & \eta^{3/5}M\,;
\ea 
are the total mass, the symmetric mass ratio and the ``chirp" mass, respectively, of a binary of components masses $m_1$ and $m_2$. Lastly, $\Phi_c$ is the constant phase at some (arbitrary) reference time $t_c$.

In a theory of gravity in which the propagation velocity of gravitational radiation is different from the speed of light, $v_\mathrm{g} \neq c$ and depends on the frequency $f$ as $(v_\mathrm{g}/c)^2=1-(c/f\lambda_\mathrm{g})^2$, there is a very simple imprint on the phase of the radiation, as shown by Will in Ref.~\cite{Will:1998}. This is equivalent to the effect that a graviton (the particle mediating the gravitational interactions) with non-zero mass (or a finite size graviton Compton wavelength) would induce. Following the accepted convention in the literature, we will can this a ``massive-graviton (MG) theory". In this paper, we will consider the result derived in Ref.~\cite{Will:1998} for the expression of the waveform emitted by a binary system in a MG theory:
\be
h_\mathrm{MG}(f)={\cal A}f^{-7/6}e^{i\Psi_\mathrm{MG}(f)}\,.
\label{e:hMG}
\ee
The amplitude is the same as for the General Relativity case, Eq.~(\ref{A:gw}), and the gravitational-wave phase becomes
\be
\Psi_\mathrm{MG}(f) = \Psi_\mathrm{GR}(f) - \frac{\pi^2 D M}{\lambda_\mathrm{g}^2 (1+z)}(\pi M f)^{-1}\,,
\label{e:phaseMG}
\ee
where $z$ is the cosmological redshift of the source. It is useful to notice that the frequency dependency of the massive-graviton term is proportional to $(\pi M f)^{-1}$, which is the same as the first post-Newtonian contribution, see Eq.~(\ref{e:phaseGR}).
The distance $D$ that appears in the massive-graviton phase term~(\ref{e:phaseMG}) is given by:
\be
D \equiv \frac{1+z}{H_0}\int_0^{z}\frac{dz'}{(1+z')\sqrt{\Omega_M (1+z'^3)+\Omega_{\Lambda}}}\,,
\label{distance}
\ee
where we have assumed a flat universe, and $H_0$, $\Omega_M$ and $\Omega_{\Lambda}$ are the Hubble parameter, the matter and cosmological constant density parameters, today; in this paper we will use values consistent with the standard $\Lambda$CDM cosmology, $H_0 = 70$ km s$^{-1}$ Mpc$^{-1}$, $\Omega_M = 0.3$ and $\Omega_{\Lambda} = 0.7$~\cite{KomatsuEtAl:2010}. We note that $D$ in general differs from the luminosity distance $D_L$ that enters into the GW amplitude, Eq.~(\ref{A:gw}), by a factor $(1=z)^{-1}$. However, in this paper we concentrate on sources within $150$ Mpc, $z < 0.04$ observed with ground-based laser interferometers with sensitivity typical of second generation. With these assumptions, $D$ differs from $D_L$ by less than $5\%$ and we set the two numbers to be the same, and ignore the difference, and therefore present the results directly as a function of $\lambda_\mathrm{g}$. In real observations, $z$ cannot be measured from gravitational wave observations alone, and therefore one cannot measure independently $D$ and $\lambda_\mathrm{g}$: one can only refer to the combination $\propto D/[\lambda_\mathrm{g}^2 (1+z)]$. None of these assumptions have an impact on the conceptual approach that we propose in this paper. However, some of the results on the specific cases considered here -- an alternative theory characterised a massive graviton -- are affected by the approximation. During the actual analysis of the data one would use the most accurate expression of the waveform available.

\subsection{Hypotheses}

In this paper we use consider tests of General Relativity versus a massive-graviton theory, according to the assumptions that we have discussed in the previous Section. Here we explicitly state the models (or, equivalently, hypotheses) that we consider and are codified with the following notation:
\begin{itemize}
\item $\mathcal{H}_\mathrm{GR}$: The data consists of (zero mean) Gaussian and stationary noise of known spectral density plus an inspiral signal of the form described by Eqs.~(\ref{e:hGR})-(\ref{e:phaseGR}) with $\lambda_\mathrm{g}^{-1}=0$.
\item $\mathcal{H}_\mathrm{MG}$: The data consists of (zero mean) Gaussian and stationary noise of known spectral density plus an inspiral signal of the form described by Eqs.~(\ref{e:hMG})-(\ref{e:phaseMG}), but with $\lambda_\mathrm{g}$ as an additional unknown free parameter.
\end{itemize}
In our analysis, for both models $\mathcal{H}_\mathrm{GR}$ and $\mathcal{H}_\mathrm{MG}$ we actually compute the Bayes factors between the hypothesis that the data contain noise and signal, according to a given theory, versus the hypothesis that the data contain only only noise. We call these Bayes factors $B_\mathrm{GR,noise}$ $B_\mathrm{MG,noise}$, for a signal vs noise within General Relativity and a massive-graviton theory, respectively (see Section II, and in particular II.D of Ref.~\cite{VeitchVecchio:2010} for more details). The Bayes factor between the MG and the GR models $B_\mathrm{MG,GR}$ is then calculated simply as 
\begin{equation}
B_\mathrm{MG,GR}=\frac{B_\mathrm{MG,noise}}{B_\mathrm{GR,noise}}\,,
\label{e:BMGGR}
\end{equation}
see Eqs.~(\ref{e:bayes}) and~(\ref{e:marglikelihood}).

\section{Results}
\label{s:results-single}

In this Section we present the results pertinent to the analysis of simulated gravitational wave observations using individual detections. In Section~\ref{s:results-combined} we shall then generalise our approach to \emph{the combination of multiple gravitational-wave events} to produce more constraining statements on alternative theories of gravity, and in the specific case considered in this paper a massive graviton. In this exemplificative study, we shall use as reference sensitivity the one of second generation ground-based instruments, such as Advanced LIGO~\cite{noise}. 

We begin in Section \ref{ss:mg-inj} by discussing how to distinguish between the MG and GR models using Bayesian model selection. We consider data sets in which coalescing binary waveforms are present and are generated using MG waveforms -- we inject these waveforms into Gaussian and stationary noise for a fixed massive graviton wavelength and binary system parameters -- and analyse the data using both MG and GR waveforms to calculate the value of the Bayes factor $B_\mathrm{MG,GR}$, Eq.~(\ref{e:BMGGR}). In Section \ref{ss:bias} we then discuss the effect of analysing the data with GR waveforms -- focussing specifically on the issue of parameter estimation -- in the case in which the actual correct theory of gravity is represented by a massive-graviton theory; we therefore address the issue of bias (in particular) as well as statistical errors in parameter estimation which, except the pseudo-analytical approach
in Ref.\cite{CutlerVallisneri:2007}, cannot be addressed (and has not been addressed) solely with Fisher Information Matrix calculations. Finally, as the value of the graviton wavelength $\lambda_\mathrm{g}$ is already  tightly bound by Solar System tests and measurements of galaxy motions in clusters of galaxies~\cite{Will:2006}, we consider the case in which the correct theory of gravity is General Relativity ($\lambda_\mathrm{g} \rightarrow \infty$) and investigate how to set bounds on $\lambda_\mathrm{g}$ and how tight these constraints can be expected to be. In order to do so, we inject GR signals and analyse the data sets using massive-graviton waveforms to compute bounds on $\lambda_\mathrm{g}$ from the marginalised PDF of $\lambda_\mathrm{g}$ that the analysis yields.

Before discussing the results in Sections~\ref{ss:mg-inj}-\ref{ss:bounds} and Section~\ref{s:results-combined} we provide details about the simulations on which our analyses are based. 

\subsection{Details of the simulations}
\label{ss:overview}

Our results are derived from a set of simulations on synthetic data sets analysed using a nested sampling algorithm that we have developed for ground-based observations of coalescing binaries~\cite{VeitchVecchio:2010}. Here we provide details on the parameters of the simulations that are used throughout the paper. 

We generate Gaussian and stationary noise in the frequency domain with noise spectral density representative of second-generation instruments~\cite{noise} and superimpose (``inject", in the jargon of the LIGO Scientific Collaboration) frequency domain gravitational-wave inspiral signals corresponding to either the General Relativity model, Eqs.~(\ref{e:hGR})-(\ref{e:phaseGR}), or the massive graviton model, Eqs.~(\ref{e:hMG}) and~(\ref{e:phaseMG}). Observations to gain information on the graviton dispersion relation require the use of at least three instruments at geographically separated locations to fully break the distance-source position determination, see Eq.~(\ref{e:phaseMG}). However, in order to reduce the computational costs of the simulations, that are in general significant, we actually use data sets from a single instrument and assume the source sky location is known within a small area in the sky. As it is possible to recover the sky location when using a network of three or more interferometers, this constraint simulates the effect of using a network but with only one dataset to be analysed at a time and therefore has considerable advantages in terms of computational time. 

We inject inspiral signals from binary systems with chirp mass $\Mc = 5 \Ms$ and symmetric mass ratio $\eta=0.15$ -- the individual component masses are $m_1 = 2.9M_{\odot}$ and $m_2 = 12.7M_{\odot}$ -- and fixed, but random polarisation, inclination and sky location. In order to explore how the results depend on the signal-to-noise ratio (SNR), we repeat the injections considering different distances.  Note that if one increases the distance to the source, the gravitational waveforms is affected in two ways: the amplitude decreases (and so does the SNR), and the contribution of the massive-graviton term to the phase increases, see Eq.~(\ref{e:phaseMG}). Unless otherwise stated, we generated MG waveforms using a value of the parameter $\lambda_\mathrm{g}=10^{15}$ m. The actual present lower bound on $\lambda_\mathrm{g}$, which is inferred from Solar System experiments, is $\lambda_\mathrm{g} \ge 2.8\times 10^{15}$ m~\cite{Will:1998}. The purpose of this section is the investigation of our method's behaviour and its feasibility as a rigorous but practical way of performing tests of GR, and so we have ignored the existing physical bound of the graviton Compton wavelength here. The magnitude of the non-GR term in Eq.~(\ref{e:phaseMG}) is in fact $\propto \lambda_\mathrm{g}^{-2}$, thus to have a significant contribution from this term we need to have small values of $\lambda_\mathrm{g}$.

In the analysis, the prior distributions on the parameters are chosen as follows. We use uniform priors on the chirp mass in the range $1 \leq \Mc/\Ms \leq 15$, in $\eta$ over the range $0.1 - 0.25$, in $D_L$  over the range $1-1500$ Mpc. The prior on the the polarisation,  and initial phase angles are flat and cover the whole range, and is proportional to $|\sin{\iota}|$ for the inclination angle $\iota$ over the whole range too. The prior on the signal coalescence time is flat within a range $\pm 100$ msec centred on the true value. The situation where the signal position is reconstructed with a instrument network, but we are performing simulations using a single instrument is approximated by a prior on the source right ascension and declination that is flat in the angles and has a width of 0.1 radians around the actual value of the injection. 

When we analyse the data considering the MG model, one of the parameters of the signal is of course $\lambda_\mathrm{g}$. The functional form that we choose for the prior on $\lambda_\mathrm{g}$ is the scale invariant prior 
\be
p(\lambda_\mathrm{g}|{\cal H}_\mathrm{MG}) \propto 
\begin{cases}
1/\lambda_\mathrm{g} & \mbox{if } \lambda^\mathrm{(min)}_\mathrm{g} \le \lambda_\mathrm{g} \le \lambda^\mathrm{(max)}_\mathrm{g},\\
0 & \mbox{elsewhere.}
\end{cases}
\label{lambda:prior}
\ee
We choose the lower bound of the prior range as $\lambda^\mathrm{(min)}_\mathrm{g} = 10^{14.5}\,\mathrm{m}$ and we analyse the data using different  upper-limits in the range $10^{15.5}\,\mathrm{m} \le \lambda_\mathrm{g} \le 10^{20.5}\,\mathrm{m}$. The prior (\ref{lambda:prior}) is effectively non-informative on the scale of $\lambda_\mathrm{g}$ between its bounds~\cite{jeffrey61}, and also implies that $p(\log{}\lambda_\mathrm{g}|{\cal H}_\mathrm{MG})=$ const. In the numerical integration, we actually consider $\log{}\lambda_\mathrm{g}$ as massive-graviton parameter as it is more easily sampled over this large a range.

%
%
\begin{figure*}[t]
\includegraphics[width=6.5in]{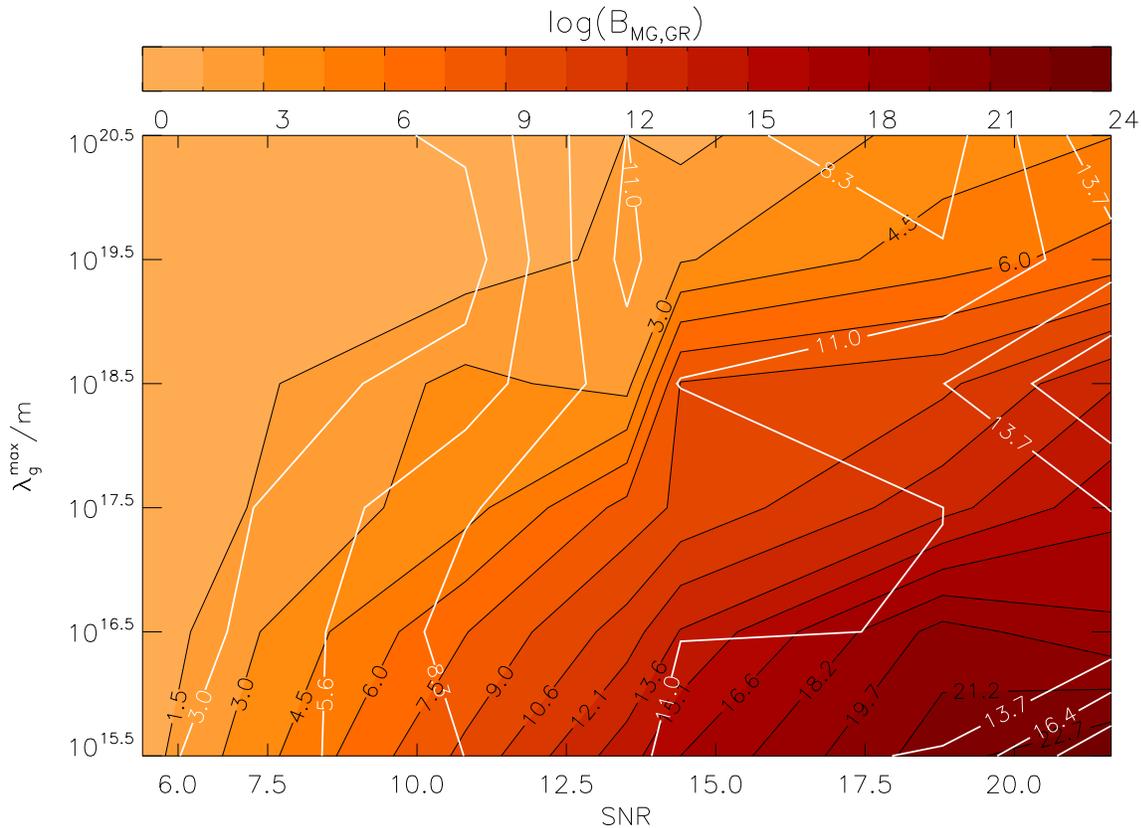}
\caption{The Bayes factor of the massive-graviton vs General Relativity 
model $B_\mathrm{MG,GR}$, see Eq.~(\ref{e:BMGGR}) obtained by analysing 
data sets containing a massive-graviton waveform as a function of the 
signal-to-noise ratios (SNR) of the injection and the prior range on 
$\lambda_\mathrm{g}$. For each value of the injection and analysis 
parameters, 10 data sets with independent noise realisations are 
generated and processed in order to quantify the variation of 
$B_\mathrm{MG,GR}$ produced by the stochastic nature of the noise. The 
black contours show the median value, while the white contours show 
the $95\%$ confidence interval of $\log{}B_\mathrm{MG,GR}$. For all the injections, the mass parameters are $\Mc=5\Ms$, $\eta = 0.15$ and $\lambda_\mathrm{g} = 10^{15}$m. The location and orientation of the source in the sky were drawn randomly and kept fixed in each run, and the distance changed in order to produce different signal-to-noise ratios. The lower-bound of the prior massive-graviton wavelength is kept constant in all the analyses and set to $\log_{10}(\lambda^\mathrm{(min)}_\mathrm{g}/\mathrm{m}) = 14.5$; as upper-bound we consider in turn six values (spaced by an order of magnitude), from $\log_{10}(\lambda^\mathrm{(max)}_\mathrm{g}/\mathrm{m}) = 15.5$ to 20.5 and are shown on the vertical axis. The Bayes factor decreases with the increase of the prior width because the evidence, Eq.(\ref{e:marglikelihood}), is smaller when integrated over a larger volume in the parameter space.}
\label{f:bayes}
\end{figure*}
%
%

The analysis of the simulated data sets is carried out using the nested sampling algorithm that we have developed for ground-based observations of coalescing binary systems, with a total of 1\,000 ``live points'' for each analysis. Technical details about the method are available in Ref.~\cite{VeitchVecchio:2010} and the software is released as part of the \texttt{ lalapps\_inspnest } program, which can be found in the LIGO Scientific Collaboration LALApps software distribution~\cite{lalapps}. 

The stochastic nature of the noise affects the actual value of the Bayes factor that is recovered by the analysis, which varies from noise realisation to noise realisation for fixed values of the signal injection parameters. In several instances, in the results that we provide below we average over the noise realization, that is we inject the same signal on many  different and statistically independent noise realisations and we report (some measure of) the spread of the Bayes factors. We also note that we have chosen the tunable parameters of the nested sampling algorithm used in the analysis such that the fluctuations due to the noise dominate any fluctuation of the Bayes factor values due to the Monte Carlo nature of the technique.

\subsection{Selecting a theory}
\label{ss:mg-inj}

We first consider the problem of discriminating between a massive-graviton theory and General Relativity. In order to do so, we generate signals from binary systems with mass parameters $\Mc = 5 \Ms$ and $\eta = 0.15$, as we have described in the previous Section. We fix the value of the Compton wavelength to $\lambda_\mathrm{g}=10^{15}$m, and in order to explore the dependency of the results on the signal-to-noise ratio (SNR) we repeat the injections using a different value of the luminosity distances, to span the range $5 \simlt \mathrm{SNR} \simlt 20$. We also generate and analyse 10 data sets with independent noise realisations for fixed injection parameters. In the analyses, the prior on $\lambda_\mathrm{g}$ is given by Eq.~(\ref{lambda:prior}); however, for each data set the analysis is repeated for six different ranges of $p(\lambda_\mathrm{g}|{\cal H}_\mathrm{MG})$, with $\lambda^\mathrm{(min)}_\mathrm{g} = 10^{14.5}$m (which is kept constant), and $\lambda^\mathrm{(max)}_\mathrm{g}$ is changed by an order of magnitude each time to span the range $\lambda^\mathrm{(max)}_\mathrm{g} = 10^{15.5}\,\mathrm{m} - 10^{20.5}$m.

The results of the analysis are summarised in Figure~ \ref{f:bayes}, 
where we plot the Bayes factors $B_\mathrm{MG,GR}$, see 
Eq.~(\ref{e:bayes}), as a function of the optimal SNR of the injection 
and the upper bound of the prior on the Compton wavelength. For each 
value of the prior range and the SNR, $B_\mathrm{MG,GR}$ shows a range 
of values (in fact we plot  $\mathrm{max}(B_\mathrm{MG,GR})$ and 
$\mathrm{min}(B_\mathrm{MG,GR})$, the maximum and minimum Bayes factor 
that are computed, respectively, due to the effect of the different 
realisation of the noise. The results indicate that, for any given 
prior range on $\lambda_\mathrm{g}$, the Bayes factors 
$B_\mathrm{MG,GR}$ is, on average, a monotonic increasing function of the SNR. At the same time, for a given SNR, $B_\mathrm{MG,GR}$ is a monotonic \emph{decreasing} function of the prior range on $\lambda_\mathrm{g}$. To understand this last behaviour, consider that the integral in Eq.(\ref{e:marglikelihood}) is performed over the whole parameter space. For a given likelihood, the effect of enlarging the prior on $\lambda_\mathrm{g}$ is then to decrease the evidence. Nevertheless, we note that, in our simulations, $\log{}B_\mathrm{MG,GR}$ favours the MG model for SNR $\simgt 15$ and for priors on $\lambda_\mathrm{g}$ that span at most two orders of magnitude. 

%
%
\begin{figure*}[t]
\begin{tabular}{cc}
\resizebox{\columnwidth}{!}{\includegraphics{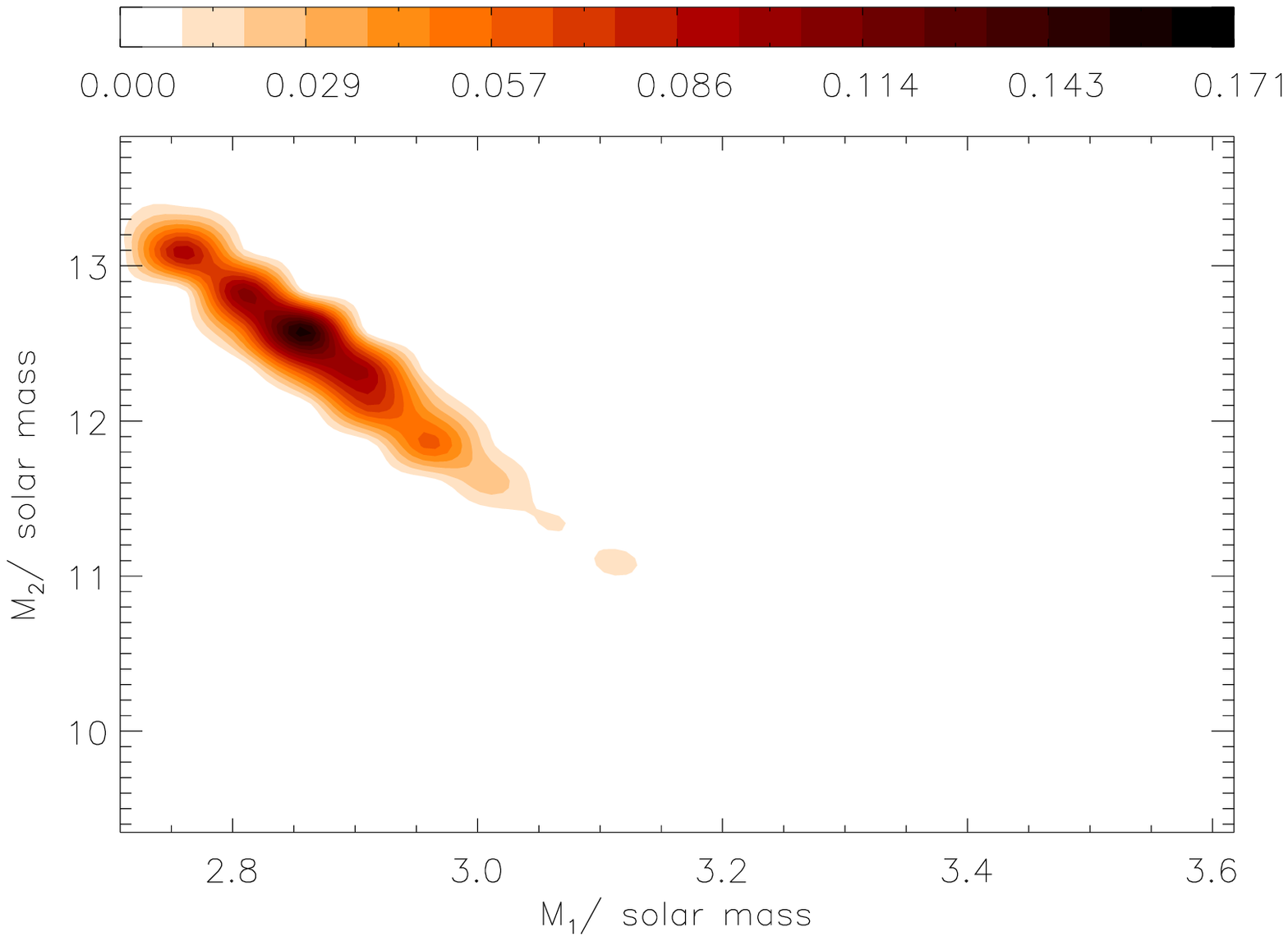}} & \resizebox{\columnwidth}{!}{\includegraphics{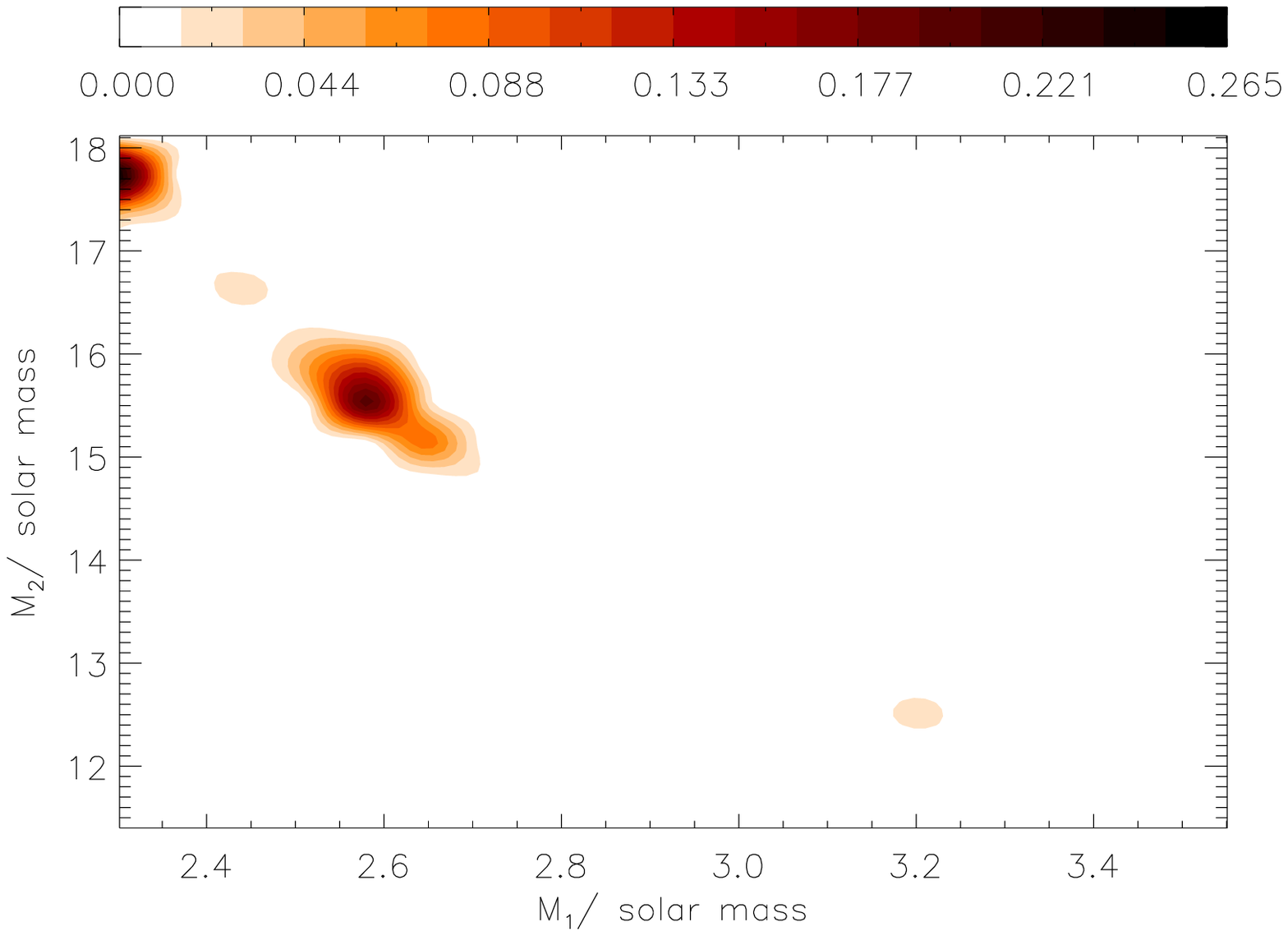}} \\
\resizebox{\columnwidth}{!}{\includegraphics{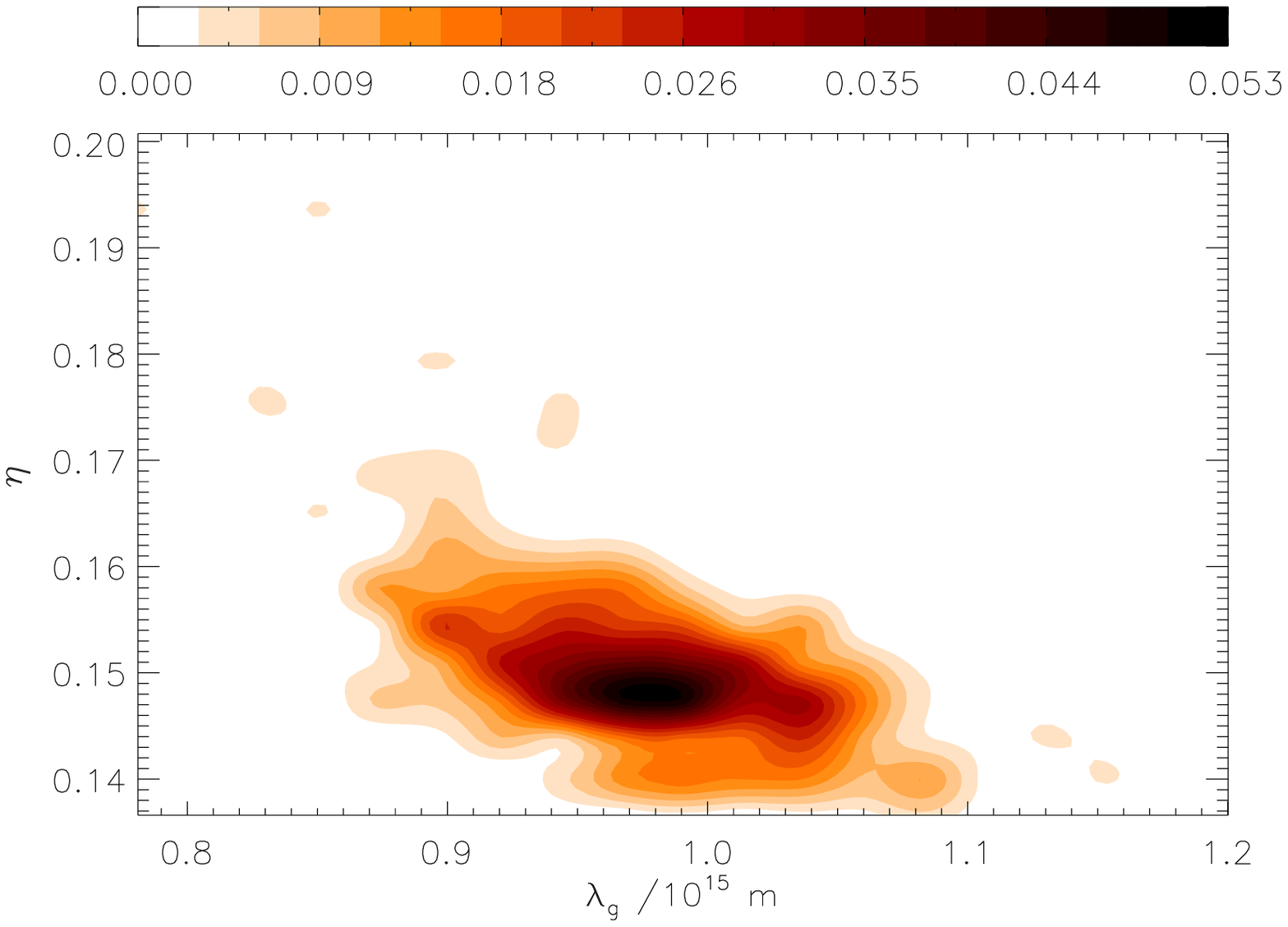}} & \resizebox{\columnwidth}{!}{\includegraphics{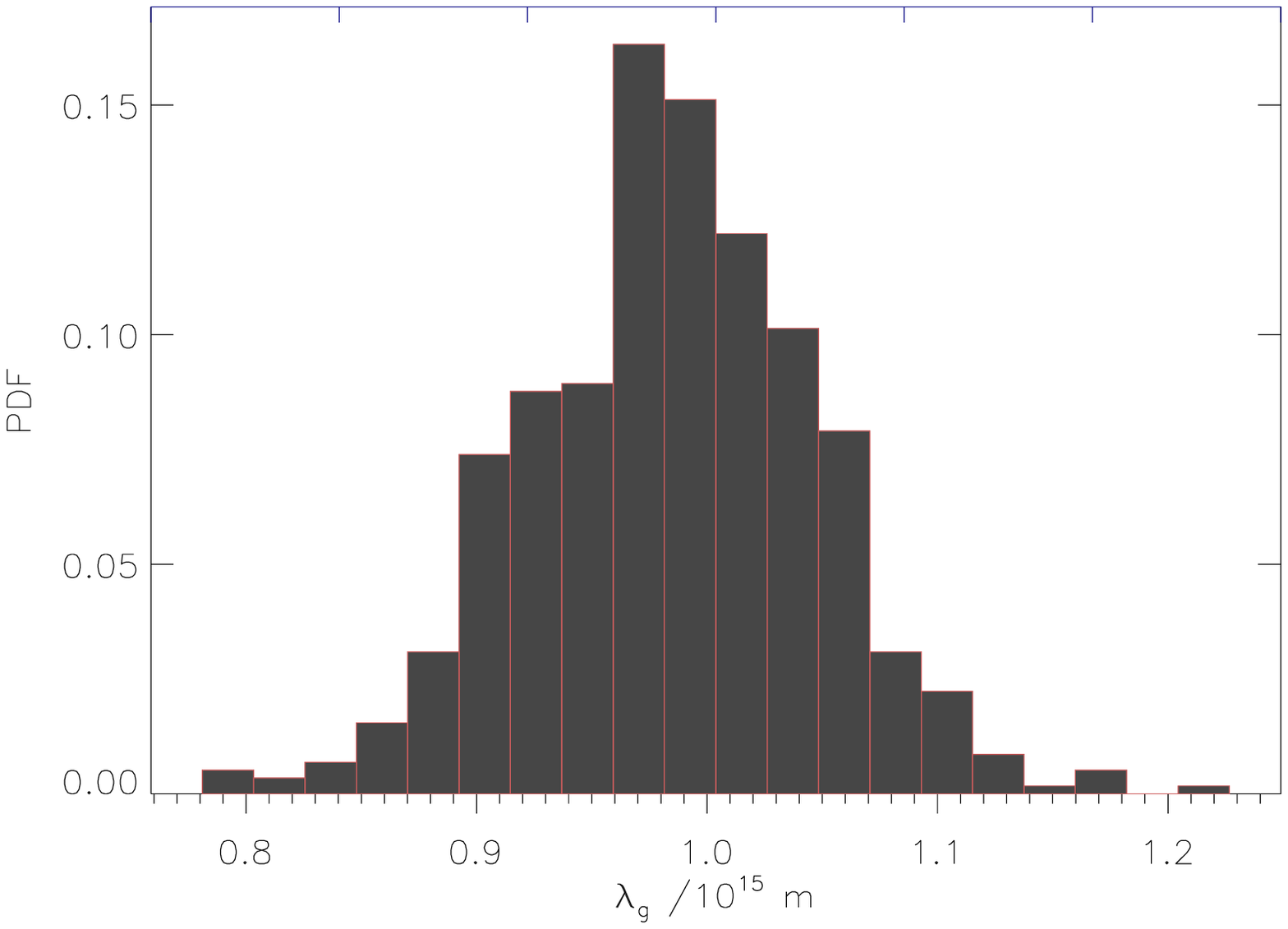}}  \\
\end{tabular}
\caption{Marginalised posterior density functions of selected parameters obtained by analysing a data set in which a massive-graviton signal was injected with parameters $\Mc = 5.0 M_{\odot}$, $\eta = 0.15$ (giving $m_1=2.9 \Ms$ and $m_2=12.7 \Ms$) and $\lambda_\mathrm{g}= 10^{15}$m at a distance of $D_L=60$Mpc. Sky position and orientation of the orbital angular momentum are randomly chosen and yield an optimal SNR of $14.4$. The corresponding number of wave cycles in the sensitivity band associated to the massive graviton term is 27. The prior on the massive graviton is flat in $\log_{10}(\lambda_\mathrm{g})$, and non-zero for $\log_{10}(\lambda_\mathrm{g}/\mathrm{m}) \in [14.5,20.5]$. The analyses yield $\log B_\mathrm{MG,noise} = 83$, $\log B_\mathrm{GR,noise} = 65$ and therefore $\log B_\mathrm{MG,GR} = 18$. Clearly the signal hypothesis is preferred to the noise-only hypothesis using both GR and MG waveforms. However, the Bayes factor is clearly in favor of the MG model rather than the GR model. \emph{Top left panel}: The marginalised posterior PDFs of $m_1$ and $m_2$ obtained by analysis the data with MG model, $p(m_1,m_2 | d, \mathcal{H}_\mathrm{MG})$. Note that the distribution is peaked at the true value of the injection parameters. \emph{Top right panel}: The marginalised posterior PDF of $m_1$ and $m_2$ obtained by analysis the data with GR model, $p(m_1,m_2 | d, \mathcal{H}_\mathrm{GR})$. The posterior PDF now is offset from the true value of the parameters and shows a bias in the recovery of the parameters. \emph{Bottom left panel}: The marginalised posterior PDF of $\eta$ and $\lambda_\mathrm{g}$ obtained by analysis the data with the MG model, $p(\eta,\lambda_\mathrm{g} | d, \mathcal{H}_\mathrm{MG})$. \emph{Bottom right panel}: the marginalised posterior distribution of $\lambda_\mathrm{g}$ using the MG model, $p(\lambda_\mathrm{g} | d, \mathcal{H}_\mathrm{MG})$. When the data are analysed with the MG model (which corresponds to the injected waveform model) the signal parameters are correctly recovered.}
\label{f:parameters}
\end{figure*}
%
%

A byproduct of our analysis are the marginalised PDFs, see 
Equation~(\ref{e:pdf}), of the parameters characterising the relevant 
model. Figure \ref{f:parameters} shows the marginalised PDFs for selected parameters for a specific injection (and noise realisation) with SNR $= 14.4$. For such system the total number of wave cycles produced by the massive graviton term in Eq.~(\ref{e:phaseMG}) within the sensitivity band is $27$. The prior range on $\lambda_\mathrm{g}$ is $10^{14.5} \mbox{m} \le \lambda_\mathrm{g} \le 10^{17.5} \mbox{m}$. For the specific noise realisation, we obtained $\log B_\mathrm{MG,noise} = 81.7$ and the $\log B_\mathrm{GR,noise} = 65$ giving $\log B_\mathrm{MG,GR} = 18$. The signal is clearly detected using both the MG and the GR model, however the MG model is favored over GR by $\approx 65,\,000,\,000$ to 1. Using the MG model, the parameters are correctly recovered, which is not the case for the GR model, where a significant bias is observed (see top-right panel of Figure \ref{f:parameters}). This is an issue that has not been considered so far by other studies and will be discussed in detail in Section~\ref{ss:bias}.

%
%
\begin{figure}[t]
\resizebox{\columnwidth}{!}{\includegraphics{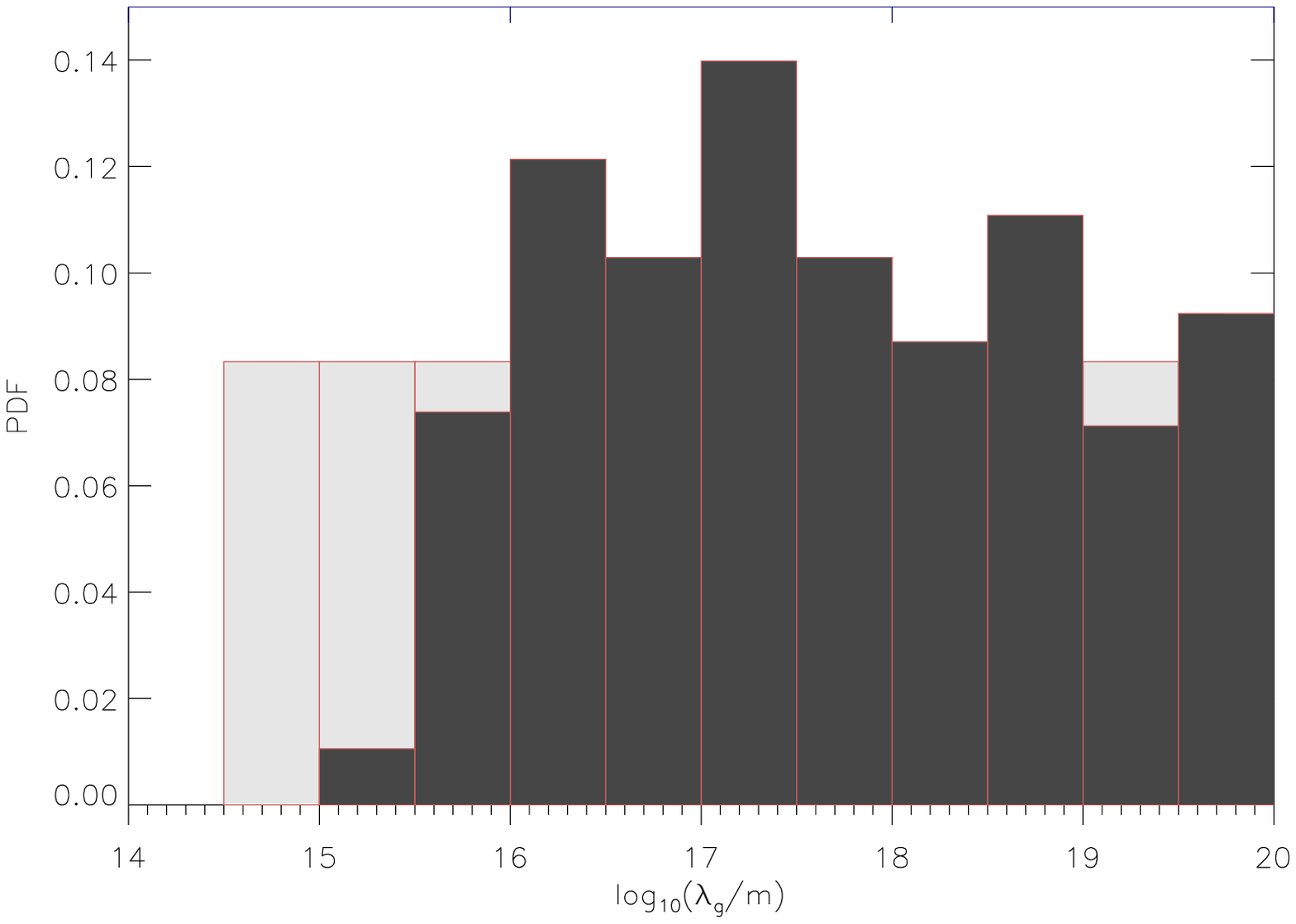}}
\caption{The prior and marginalised posterior probability density 
function of $\log_{10}(\lambda_\mathrm{g})$ for the analysis of the 
same injection parameters used for Figure~\ref{f:parameters}. In this analysis however the prior distribution is non-zero over the interval $14.5 \le \log_{10}(\lambda_\mathrm{g}/\mathrm{m}) \le 20.5$. For this particular noise realisation $\log B_\mathrm{MG,noise}=82$ while $\log B_\mathrm{GR,noise}=81$. The charcoal coloured histogram is an example PDF for $\log_{10}(\lambda_\mathrm{g})$ when $\log B_\mathrm{MG,GR}$ does not favor the MG model because of the prior width. The injected waveform has $\Mc = 5 \Ms$, $\eta=0.15$ and $\lambda_\mathrm{g}=10^{15}$ m, with optimal SNR of $14.4$. The light grey histogram is the prior distribution of $\log_{10}(\lambda_\mathrm{g})$.}
\label{f:lambda-pdf}
\end{figure}
%
%

When the width of the prior on $\lambda_\mathrm{g}$ is large enough to dominate the integral (\ref{e:marglikelihood}) and therefore the Bayes factor does not favour the MG model, the posterior density function on $\lambda_\mathrm{g}$ \emph{if one assumes the MG model} shows a characteristic behavior, which is shown in Figure~\ref{f:lambda-pdf}: the analysis excludes the region of the $\log_{10}(\lambda_\mathrm{g})$ prior that would give raise to an appreciable effect, ruling out the lower parts of the prior distribution. 

\subsection{Parameter estimation and bias}
\label{ss:bias}

The assumption that General Relativity is the correct theory of gravity may affect parameter estimation (and therefore inference) if the actual theory is different. The impact is clearly not only restricted to the statistical errors with which the parameters are recovered but may also lead to bias. One might think that if a non-GR signal is detected with GR templates, then the effects on the estimation of the parameters is negligible. However, this is not necessarily true, as we have already shown in Figure~\ref{f:parameters} (top-right panel): the in-spiral binary signal is clearly detected ($\log B_\mathrm{GR,noise} = 65$), but the mass estimates are unambiguously biased; in fact the signal present in the data follows the massive graviton model.

The particular aspect of bias is very important, as it introduces a conceptually different source of error -- a systematic offset from the true value which does not fade away as the signal-to-noise ratio increases  -- that has completely been ignored so far. In fact, all the work on so-called parameter estimation in the context of other theories of gravity has been carried out so far using the Fisher information matrix formalism that \emph{assumes} that there is no bias and deals only with statistical errors. The result that we have just highlighted in Figure~\ref{f:parameters} shows that this is a key issue. Using the MG model as an example of deviation from General Relativity, we show in this section the possible effects. Of course every theory of gravity would lead to different results; in the future, it is therefore important to study each one of them individually. For the specific case of a massive graviton theory, we note that the phase contribution of $\lambda_\mathrm{g}$ is proportional to $(\pi \Mc f)^{-1}$, see Eq.~(\ref{e:phaseMG}): it scales as the General Relativity post$^{1}$-Newtonian contribution, which is controlled by $\eta$. Therefore we expect that the parameter that will be mostly affected by processing the data with a GR waveform will be the symmetric mass ratio $\eta$ (and in turn the estimate of the individual mass components). The bottom-left panel of Figure \ref{f:parameters} shows precisely the degeneracy (or correlation) between $\lambda_\mathrm{g}$ and $\eta$ (clearly in the case in which the data are analysed assuming the model MG).

In order to explore these effects, we perform a set of numerical experiments in which we inject MG waveforms according to Eq.~(\ref{e:hMG}) and (\ref{e:phaseMG}) at fixed SNR and mass parameters ($\Mc = 5 M_{\odot}$, $\eta=0.15$ and SNR = 21) and vary the value of the graviton Compton wavelength of the signal in the interval $10^{14.5}\mathrm{m} \leq \lambda_\mathrm{g} \leq 10^{17}\mathrm{m}$, using the values reported in Table~\ref{t:cycles}. For a fixed signal-to-noise ratio (or $D_L$), it is useful to notice that for $\lambda_\mathrm{g} \simgt 3.2\times 10^{15}$ m (and for the parameters of the injections) the number of wave cycles $\mathcal{N}_{MG}$ to which the massive graviton term contributes in the detection band (\emph{i.e.} for frequencies above 20 Hz, in this case) drops below 1. As usual, for each set of injection parameters we generate 10 independent noise realisations. In addition, the data are analysed three times, using three different prior ranges for $\lambda_\mathrm{g}$, see Eq.~(\ref{lambda:prior}): the lower bound of the range is kept fixed to $\lambda^\mathrm{(min)}_\mathrm{g} = 10^{14.5}$m and the upper-bound is set to $\lambda^\mathrm{(max)}_\mathrm{g} = 10^{15.5}$m, $10^{18.5}$m and $10^{20.5}$m, in turn. 

We explore the dependency of the Bayes factor $\log{}B_\mathrm{MG,GR}$ on the value of $\lambda_\mathrm{g}$ and we investigate whether in analysing the data assuming the GR model, the estimates of the unknown parameters are affected. 
%
%
\begin{table}[!ht]
\begin{tabular}{r c}            
\hline         
$\lambda_\mathrm{g}$ [m]& $\mathcal{N}_{MG}$\\
\hline
$3.16 \times 10^{14}$  &  $171$\\
$5.62 \times 10^{14}$   & $54$\\
$10^{15}$   &  $17$\\
$1.78\times 10^{15}$   & $5.4$\\
$3.16\times 10^{15}$   &  $1.7$\\
$5.62\times 10^{15}$ & $0.5$\\
$10^{16}$ & $0.2$\\
$1.78 \times 10^{16}$ & $0.05$\\
$3.16 \times 10^{16}$ & $0.02$\\
$5.62 \times 10^{16}$ & $0.005$\\
$10^{17}$ & $0.001$\\
\hline                              
\end{tabular}
\caption{Total number of wave cycles from the massive graviton phase term as a function of $\lambda_\mathrm{g}$ {for the mass parameter values used in the injections ($\Mc = 5 M_{\odot}$ and $\eta=0.15$) and a low-frequency cut-off $f_\mathrm{low} = 20$ Hz, to which results in Figure~\ref{f:bias-eta} refer.}}
\label{t:cycles}
\end{table}
%
%

\begin{figure*}[t]
\begin{tabular}{cc}
\resizebox{\columnwidth}{!}{\includegraphics{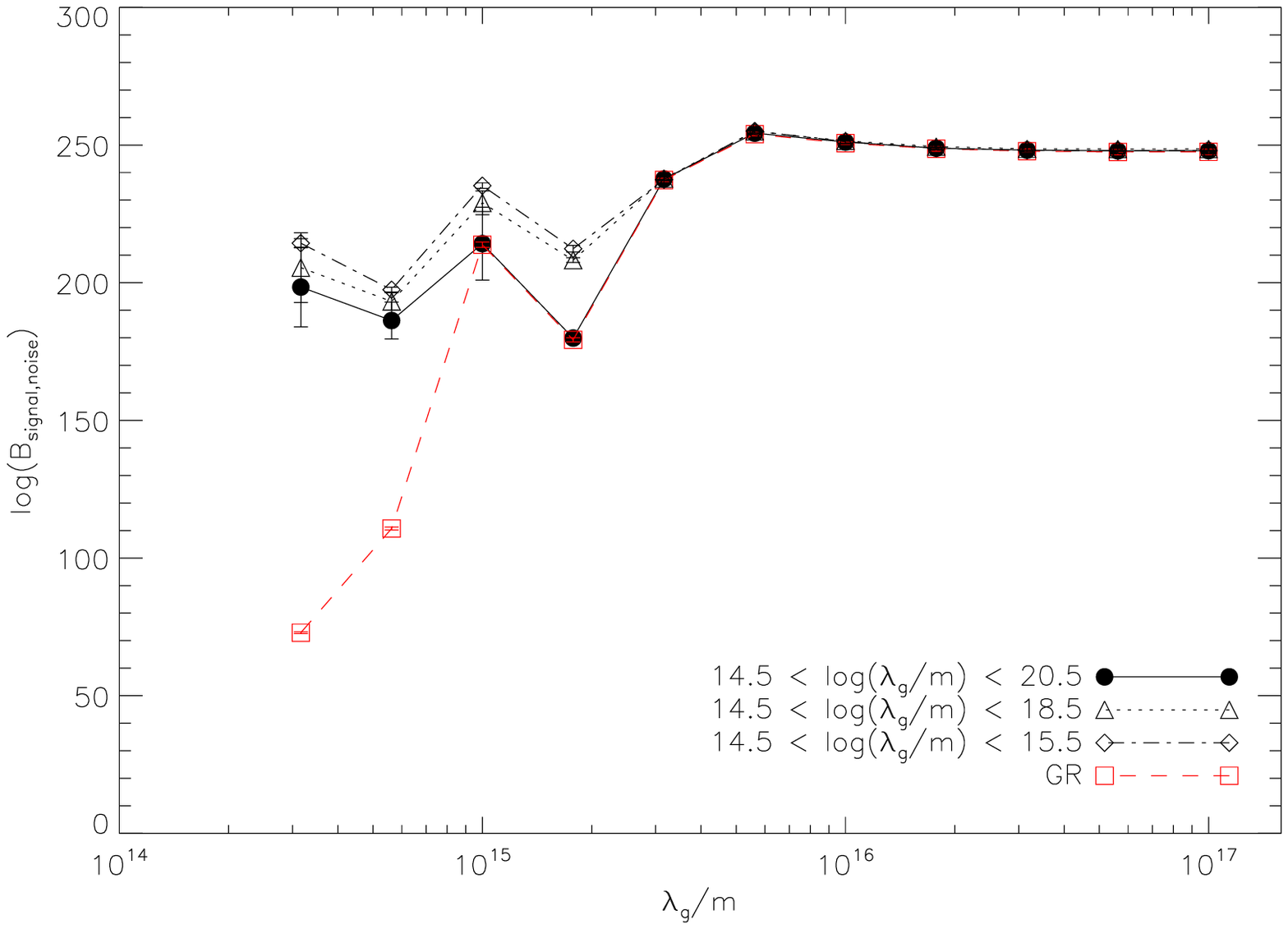}} & 
\resizebox{\columnwidth}{!}{\includegraphics{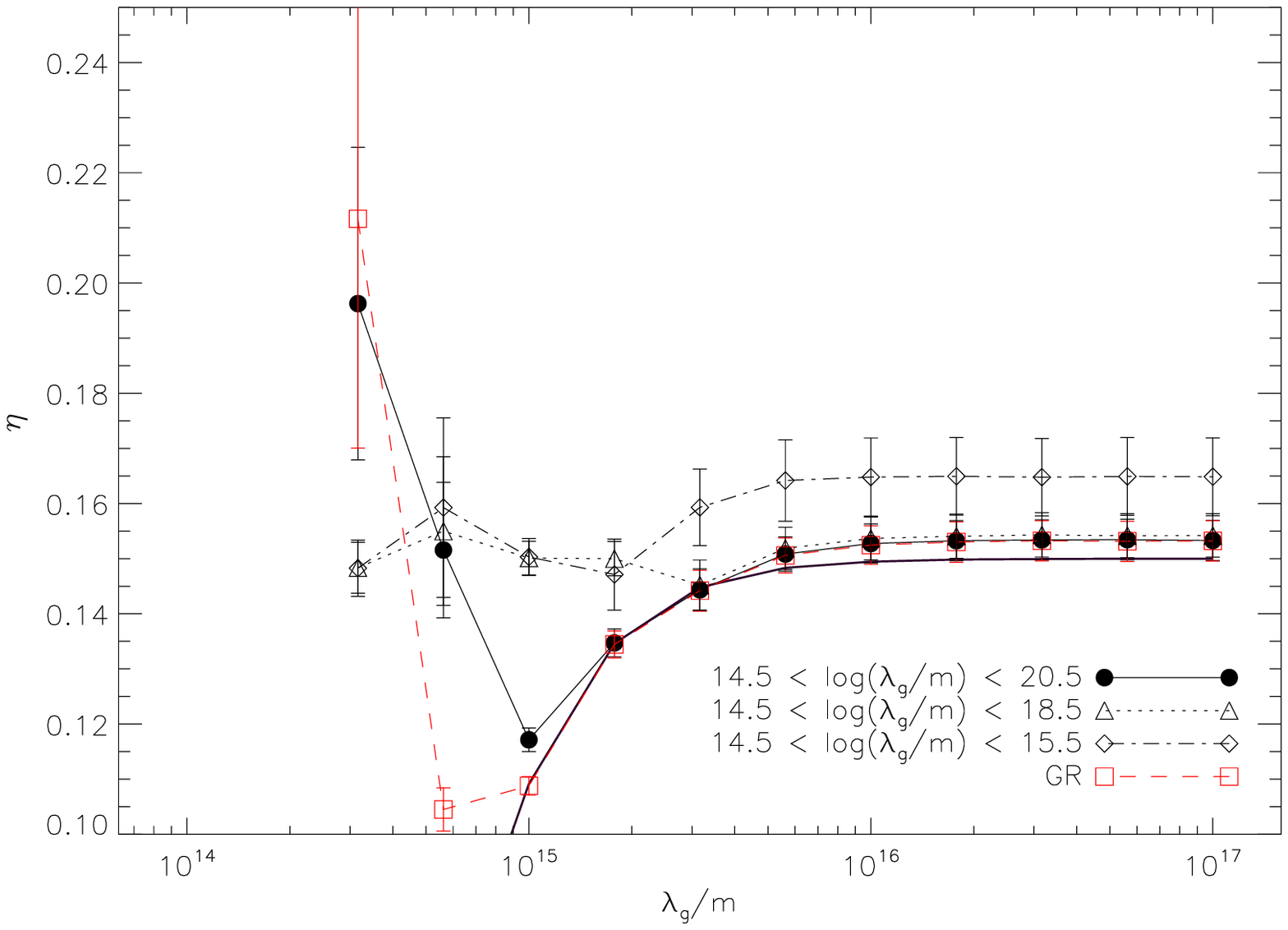}} \\
\end{tabular}
\caption{\emph{Left panel}: The Bayes factor of the signal over noise-only hypotheses assuming the MG model (black filled circles, open diamonds and open triangles) and GR (red open squares) as a function of the value of $\lambda_\mathrm{g}$ (in meters) of the injected signal. In the MG case each symbol corresponds to a different prior range on $\lambda_\mathrm{g}$ shown in the bottom-right corner of the plots. All the injections are carried out for a signal with $\Mc = 5 \Ms$ and $\eta=0.15$, and distance and angular parameters selected in such a way to produce an optimal signal-to-noise ratio of 21.6. It is clear from the fact that $\log B_\mathrm{MG,noise}$ and $\log B_\mathrm{GR,noise}$ are $\gg 1$ that the signals are unambiguously detected. As expected the MG theory is (strongly)  favoured for small values of $\lambda_\mathrm{g}$, but for $\lambda_\mathrm{g} \simgt  10^{15}$ m (for this specific choice of masses, distance and signal-to-noise ratio) $\log B_\mathrm{MG,GR} \approx 0$,  thus the data do not provide any conclusive evidence in favour of the MG theory.
\emph{Right panel}: The median of the maximum likelihood value of the symmetric mass ratio $\eta$ as recovered by the MG model and the GR model. The symbols are the same as the left panel. Each point is the result obtained by averaging over $10$ independent nested sampling realisations. The error bars represent the combination of the $95\%$ probability intervals from each run. The results described by the solid circles correspond to parameters correspond to a value of eta in the GR model that gives a number of cycles that is as close as possible to the MG (for the signal's injection parameters). Note that although $\log B_\mathrm{MG,GR} \approx 0$ for $10^{15} \simlt \lambda_\mathrm{g}/m \simlt  10^{16}$, and therefore the MG model is not favored over GR, the value of $\eta$ recovered by the GR model is systematically biased to compensate for the additional phase shift due to the mass of the graviton that the GR model can not account for properly.}
\label{f:bias-eta}
\end{figure*}

Figure \ref{f:bias-eta} summarises the results, as a function of $\lambda_\mathrm{g}$. The analyses of the data yield $\log B_\mathrm{MG,noise}$ and $\log B_\mathrm{GR,noise}$ always greater than unity and in general $\gg 1$ showing that the signal hypothesis is clearly favored over the noise hypothesis, and therefore the signal is ``detected" (left panel of Fig.\ref{f:bias-eta}). This happens regardless of the choice of the prior on $\lambda_\mathrm{g}$, and of the model (MG or GR). Furthermore, $\log B_\mathrm{MG,noise}$ is found in the interval $200 \le \log B_\mathrm{MG,noise} \le 250$, whereas $\log B_\mathrm{GR,noise}$ increases from $\approx 70$ for $\lambda_\mathrm{g} =3.16\times10^{14}$ m to the same values of $\log B_\mathrm{MG,noise}$ for $\lambda_\mathrm{g} \simgt  10^{15}$ m.  In fact, when $\lambda_\mathrm{g}$ exceeds  $\approx  10^{15}$ m (for this specific choice of masses and signal-to-noise ratio), $\log B_\mathrm{MG,GR} \approx 0$ and there is no conclusive evidence from the data that the MG model should be preferred to the GR model. This behavior is consistent with what one would expect: as $\lambda_\mathrm{g}$ increases, the MG waveform looks progressively more as a GR waveform, and this is reflected in the behavior of the Bayes factors. Eventually the Occam's Razor takes over and naturally the simplest theory is favored (if there are no strong reasons to prefer the more complex one).

The recovered median value of $\eta$ varies with the two models used to analyse the data, and within the MG model it varies with the prior chosen on $\lambda_\mathrm{g}$. For the largest prior size, $10^{14.5}\mathrm{m} \leq \lambda_\mathrm{g} \leq 10^{20.5}\mathrm{m}$, MG behaves similarly to GR (right panel of Fig.\ref{f:bias-eta}). When the size of the phase terms that are neglected in the model template is large, therefore for $\lambda_\mathrm{g} \simlt 10^{15}$ m (\emph{cf.} Table \ref{t:cycles} for our choice of parameters),  the best estimate of $\eta$ is severely biased (although the signal is clearly detected using just GR templates). Consider the contribution to the total number of cycles in the detector band from the massive graviton phase term contribution $\mathcal{N}_{MG}$. 
When $\mathcal{N}_{MG} \simlt 1$, the GR waveform does not differ significantly from a MG waveform, so small or no systematic effects are introduced in the parameter recovery. When $\mathcal{N}_{MG}  \simgt 1$, the GR template has no way to distinguish the phase shift due to the massive graviton term from the post$^{1}$-Newtonian term, see Eq.~(\ref{e:phaseGR}). Consequently, the only way in which the GR model template can match the observed MG signal is to lower the value of $\eta$ to account for the phase shift. Needless to say, even if the $\eta$ PDF is strongly peaked, the best estimate of $\eta$ is actually quite different from the injected value (notice that $\log B_\mathrm{GR,noise} \gg 1$). Without the knowledge of the ``true" model describing the observed data, even in case of a successful detection inference on the parameters may be affected. For the intermediate prior, $10^{14.5}\mathrm{m} \leq \lambda_\mathrm{g} \leq 10^{18.5}\mathrm{m}$ (empty triangles in Figure~\ref{f:bias-eta}), the MG model manages to recover the correct value of the parameter $\eta$. In this case, no bias is ever introduced. When we consider the smallest prior, $10^{14.5}\mathrm{m}  \leq \lambda_\mathrm{g}\leq 10^{15.5}\mathrm{m}$ (open diamonds in Figure~\ref{f:bias-eta}), $\eta$ is correctly estimated for $\lambda_\mathrm{g} < 2\times10^{15.5}\mathrm{m} $, corresponding approximately to the upper boundary of the prior itself. When the injected value of $\lambda_\mathrm{g}$ is greater than the upper bound of the prior probability distribution function, $\eta$ is instead overestimated. 
Our estimate is biased, but this time this is in the opposite direction compared to what we observed before.

%
%
%
%
%
%
\begin{figure*}[t]
\begin{tabular}{cc}
\resizebox{\columnwidth}{!}{\includegraphics{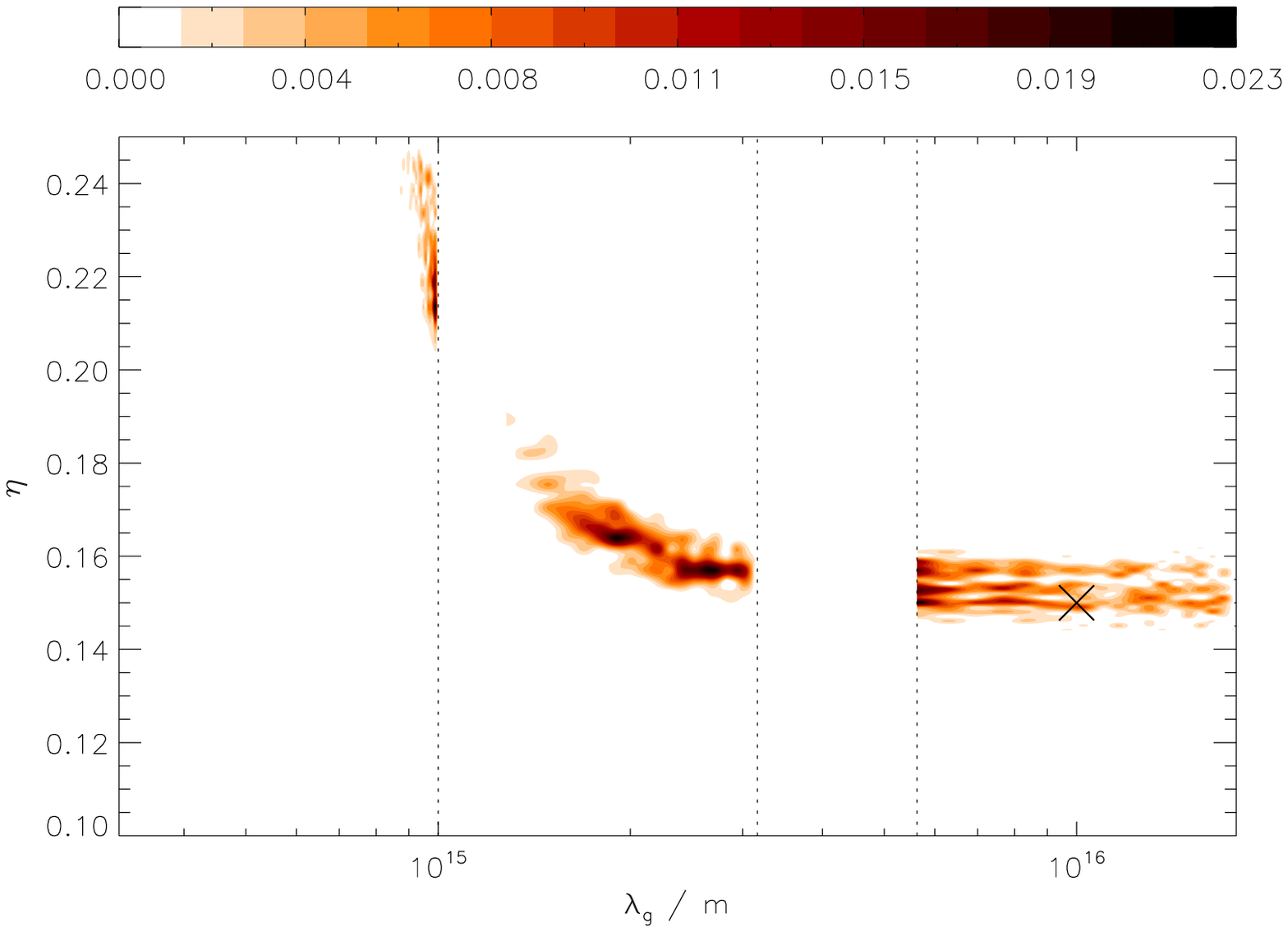}} & 
\resizebox{\columnwidth}{!}{\includegraphics{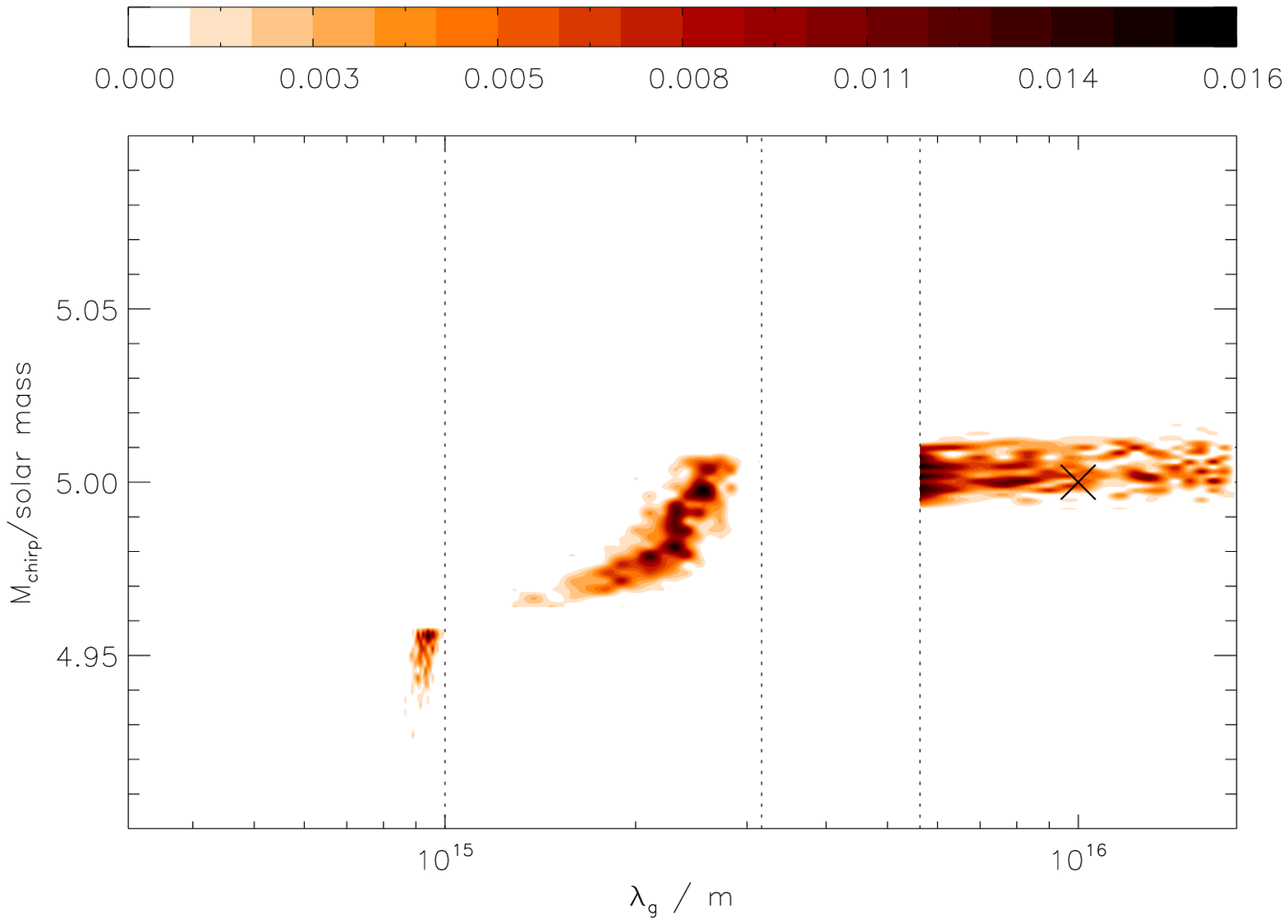}} \\
\end{tabular}
\caption{The effect of the priors on the marginalised posterior density functions of selected parameters. A massive-graviton signal was injected with $\Mc = 5 \Ms$, $\eta = 0.15$ and $\lambda_\mathrm{g}=10^{16}$m, producing an optimal SNR of 21.6. The same data set is analysed three times changing only the prior range of the graviton's Compton wavelength (everything else being the same): $10^{14.5} \le \lambda_\mathrm{g}/\mathrm{m} \le 10^{15}$, $10^{15} \le \lambda_\mathrm{g}/\mathrm{m} \le 10^{15.5}$ and $10^{15.75} \le \lambda_\mathrm{g}/\mathrm{m} \le 10^{16.25}$.  \emph{Left panel}: marginalized posterior PDF of $\eta$ and $\lambda_\mathrm{g}$. \emph{Right panel}: marginalized posterior PDF of $\Mc$ and $\lambda_\mathrm{g}$. In both panels, the vertical dotted lines show the limits of the priors used and the cross marks the injected values of the parameters. The bias in parameter estimation is clearly a function of the distance between the ``real'' (injected) value and the upper bound of the prior on $\lambda_\mathrm{g}$.}
\label{f:bias-prior}
\end{figure*}  
%
%

We further illustrate the subtleties associated to the bias in Figure~\ref{f:bias-prior}. For this we generate a single noise realisation and inject a MG waveform with $\Mc = 5 \Ms$, $\eta = 0.15$ and $\lambda_\mathrm{g}=10^{16}$m. The data set is then analysed using three different priors over the graviton Compton wavelength (same functional form as in Eq.~(\ref{lambda:prior}) but different intervals over which the prior is non-zero): $10^{14.5}\,\mathrm{m} \le \lambda_\mathrm{g} \le 10^{15}\,\mathrm{m}$, $10^{15}\,\mathrm{m} \le \lambda_\mathrm{g} \le 10^{15.5}\,\mathrm{m}$ and $10^{15.75}\,\mathrm{m} \le \lambda_\mathrm{g} \le 10^{16.25}\,\mathrm{m}$ (note that only the last interval contains the value of the injected signal). In Figure~\ref{f:bias-prior} we show the 2-dimensional marginalised posterior PDFs on the  mass parameters and the graviton Compton's wavelength. The bias on the recovered values of $\Mc$ and $\eta$ is a clear function of the distance between the injected value and the upper bound of the prior chosen to analyse the data. For $\Mc$ the recovered value differ from the real one only by few percents. $\Mc$ is in fact mainly determined by the leading Newtonian term in Eq.~(\ref{e:phaseMG}). $\eta$ and $\lambda_\mathrm{g}$ are instead anti-correlated in (\ref{e:phaseMG}), therefore when $\lambda_\mathrm{g}$ is forced to be smaller than the actual injection value, $\eta$ is pushed toward the highest acceptable values set by its own prior. It is important to notice that in the three cases, $\log B_\mathrm{MG,noise}$ is essentially the same, therefore it cannot be used to prefer one choice of the prior over another.
Nevertheless, it is possible to infer the inadequacy of a prior by checking for signs of railing against the prior boundary.

The simple examples presented in this Section are merely indicative of a possible wider problem when gravitational wave astronomy begins to take place. Severe bias in the parameter(s) recovery may be introduced by the choice of the model, \emph{cf.} Figure~\ref{f:bias-eta}, and priors, \emph{cf.} Figure~\ref{f:bias-prior}. 
%
%
\begin{figure}[!hf]
\resizebox{\columnwidth}{!}{\includegraphics{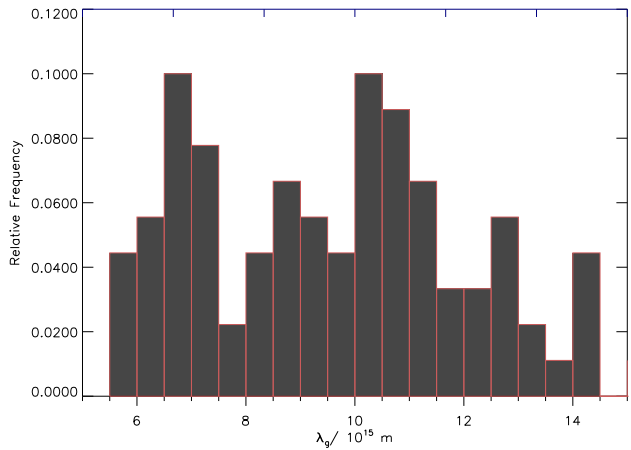}} 
\caption{The $95 \%$ lower limit on the graviton Compton wavelength $\lambda^{95 \%}_\mathrm{g}$, see Eq.~(\ref{e:95limit}), in observations with second generation ground-based instruments of inspiral signals from binaries with chirp mass $\Mc =5\,\Ms$ and $\eta = 0.15$ modeled using General Relativity waveforms. The histogram shows the relative frequency of the lower-limit from 100 injections with an optimal signal-to-noise ratio in the range $5 \le \mathrm{SNR} \le 25$. By comparison the Solar System bound on the graviton Compton wavelength is $\lambda_\mathrm{g} \ge 2.8\times 10^{15}$ m~\cite{Will:1998}.}
\label{f:95bounds}
\end{figure}
%
%

\subsection{Bounds}
\label{ss:bounds}

The results of the previous section demonstrate that if an MG theory represents the behavior of gravity, it is unlikely that in the near future gravitational wave observations will be able to provide sufficiently stringent observational results to tip the odds in favour of such a theory. On the other hand, even if the correct theory of gravity is characterised by a massless graviton, it is interesting to investigate what limit on $\lambda_\mathrm{g}$ one could place experimentally. This can be addressed in straightforward way in Bayesian inference; it simply requires the evaluation of the marginalised posterior density function $p(\lambda_\mathrm{g}|d,{\cal H}_\mathrm{MG})$, from which one can compute a lower limit on $\lambda_\mathrm{g}$ corresponding to a given probability $P$: 
\be
\int_{\lambda_\mathrm{g}^{(P)}}^{\infty} d\lambda_\mathrm{g}\, p(\lambda_\mathrm{g}|{\vec d},{\cal H}_\mathrm{MG}) = P\,.
\label{e:95limit}
\ee
In our case we decide to set (arbitrarily) $P = 0.95$ and therefore compute the $95\%$ lower limit on the graviton Compton's wavelength that we label $\lambda_\mathrm{g}^{95\%}$. In order to explore this point, we first assume that GR is the correct theory of gravity. We then produce 100 independent injections for sources with the same physical parameters ($\Mc = 5\,\Ms$ and $\eta = 0.15$) and luminosity distance but different location/orientation in the sky (drawn uniformly on the two-sphere) so as to produce a range of optimal signal-to-noise ratios. The inspiral waveform used to generate the injection is the GR waveform described by Eqs.~(\ref{e:phaseGR}) and (\ref{e:hGR}), and we analysed the data using the Massive Graviton model, Eqs.~(\ref{e:hMG}) and (\ref{e:phaseMG}) with a prior $p(\lambda_\mathrm{g}|{\cal H}_\mathrm{MG})$  with $10^{14.5}\,\mathrm{m} \le \lambda_\mathrm{g}\le 10^{20.5}\,\mathrm{m}$. From the marginalised PDF of $\lambda_\mathrm{g}$ we  compute the $95\%$ lower limit, by setting $P = 0.95$ in Eq.~(\ref{e:95limit}). The results are shown in Figure \ref{f:95bounds}. Advanced LIGO could therefore put a tighter constraint on the value of $\lambda_\mathrm{g}$ than the one currently inferred from the observation of the orbit of Mars. 
Also, Bayesian model selection allows slightly tighter bounds on $\lambda_\mathrm{g}$ compared to Fisher Information Matrix based studies. For values of the parameters similar to the ones we used, the typical lower limit from second generation instruments such as Advanced LIGO is of the order of $\mathrm{few}\times 10^{15}$ m \cite{Will:1998,ArunWill:2009}.  

\section{Combining multiple detections}
\label{s:results-combined}

Beyond the present challenge of directly detecting gravitational waves for the first time, second and third generation instruments are expected to detect an increasing number of signals in the coming years~\cite{lvc-cbc-rates,CutlerThorne:2002,Kokkotas:2008}. Therefore, it is imperative that we exploit the information that multiple observations can bring in a statistical way. Bayes' theorem offers the possibility of combining the results from each individual observation in a conceptually simple way. In the context of testing theories of gravity, this may be particularly powerful if a given theory is characterised by some ``global parameters'' -- \emph{e.g.} the Compton wavelength of gravitons -- that are independent of the actual gravitational wave signal at hand. In this case one can construct posterior density functions that take into account all the data available and therefore strengthen the inference process. For the specific case considered in this paper, we will consider how one can set more stringent lower limits on the graviton's Compton wavelength using observations of a number of coalescing binaries each of which with different parameters. Specifically in our case we consider the example of inferring $\lambda_\mathrm{g}$ from the combined probability distribution from multiple, independent, observations. 

Let us assume that we have a set of $N$ independent observations,  $d_1,\dots,d_N$, in which a gravitational wave signal from a coalescing binary is detected (here $N$ should not be confused with number of dimensions of the signal's parameter vector $\vec{\theta}$ introduced in Section~\ref{s:method}). We want to estimate the marginal PDF of $\lambda_\mathrm{g}$ from the joint set of observations. From Bayes' theorem we can write:
\be
p(\lambda_\mathrm{g} | d_1,\dots,d_N) \propto p(\lambda_\mathrm{g}) p(d_1,\dots,d_N |\lambda_\mathrm{g} )\,,
\ee
and from the chain rule,
\be\label{combined:b}
p(d_1,\dots,d_N | \lambda_\mathrm{g}) = p(d_1|\lambda_\mathrm{g},d_2,\ldots,d_N)p(d_2,\ldots,d_N|\lambda_\mathrm{g})\,.
\ee
Since the observations are independent, Eq.~(\ref{combined:b}) simplifies to
\be
p(d_1 |\lambda_\mathrm{g},d_2 \dots,d_N ) \propto p(d_1 |\lambda_\mathrm{g})\,,
\ee
and in general we can write
\be
p(\lambda_\mathrm{g} | d_1,\dots,d_N)  \propto p(\lambda_\mathrm{g}) \prod_{i=1}^{N}p(d_i |\lambda_\mathrm{g})\,,
\label{combination}
\ee
where
\be
p(d_i |\lambda_\mathrm{g}) = \int d\vec{\theta}\, p(\vec{\theta})\, p(d_i | \vec{\theta}, \lambda_\mathrm{g})
\ee
is the marginalised likelihood for the $i$th observation. 

\subsection{A pedagogical example}
\label{s:comb-sig}

One can first develop some intuition about the benefits and power of this approach by considering a simple case, in which the relevant functions that enter the computation of $p(\lambda_\mathrm{g} | d_1,\dots,d_N)$, Eq.~(\ref{combination}), have simple analytical forms; this case is also useful to disentangle conceptual issues from practical ones related to the discrete nature of the probability density functions with which one deals in practice, and that we shall address in the next Section.

The results of Section~\ref{ss:mg-inj} show that in the case in which the correct theory of gravity is General Relativity, the posterior PDF $p(\log_{10}(\lambda_\mathrm{g}) | d)$ is well approximated by a sigmoid function, see Figure~\ref{f:lambda-pdf}. Let us therefore consider a set of $i = 1,\dots\,,N$ observations each of which yields a posterior PDF is of the form
\be
p(\log_{10}(\lambda_\mathrm{g}) | d_i) \propto \frac{1}{1+a_i e^{-b_i \log_{10}(\lambda_\mathrm{g})}}\,,
\label{e:sigmoid}
\ee
where $a_i$ and $b_i$ are real numbers that differ from observation to observation. We can simulate the outcome of $N$ observations on a range of binary systems by simply selecting the values of $a_i$ and $b_i$ and then combine the results by using Eq.~(\ref{combination}). In fact, for the specific choice of prior on $\lambda_\mathrm{g}$ that we consider here, we have
\be
p(d_i | \lambda_\mathrm{g}) \propto \frac{p(\log_{10}(\lambda_\mathrm{g}) | d_i)}{p(\lambda_\mathrm{g})} = p(\log_{10}(\lambda_\mathrm{g}) | d_i)\,
\label{e:comb1}
\ee

%
%
\begin{figure}[!ht]
\resizebox{\columnwidth}{!}{\includegraphics{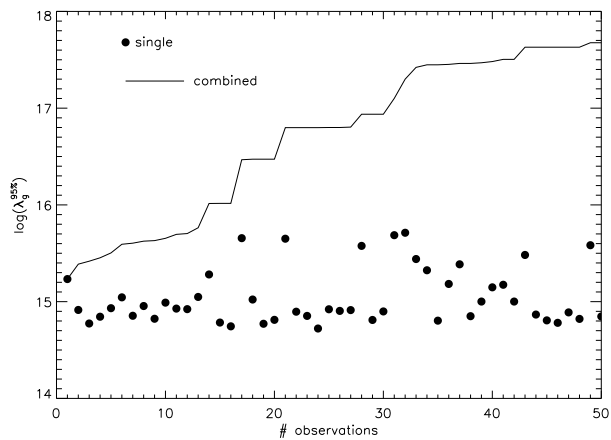}}
\caption{The 95\% lower limit on the graviton Compton wavelength $\lambda_\mathrm{g}^{95\%}$ for an example of individual and combined (independent) observations of inspiral binaries. The plot shows $\lambda_\mathrm{g}^{95\%}$  for each individual observation (solid circle) and for the combined set of observations (solid line), following Eq.~(\ref{combination}), as a function of the number of observation. The posterior PDF on $\lambda_\mathrm{g}$ from which the single and combined result are obtained is assumed to be of the form~(\ref{e:sigmoid}), with the coefficients $a$ and $b$ randomly drawn from uniform distributions in the range $a \in [2,10]$ and $b \in [1,10]$. }
\label{f:combined-sigmoid}
\end{figure}
%
%

where, differently from everywhere else in the paper, we assumed a flat prior distribution on $\lambda_\mathrm{g}$.
As an example, we can assume that we have, say, 50 detections of coalescing binaries, each of which coming from a different source and a different SNR, that leads to a different $95\%$ lower limit on the graviton's Compton wavelength $\lambda_\mathrm{g}^{95\%}$, by randomly drawing values of $a$ and $b$ to construct the PDFs~(\ref{e:sigmoid}) . Following this procedure, we have generated 50 distributions according to Eq.~(\ref{e:sigmoid}). For each of them we compute the $95\%$ lower limit on $\lambda_\mathrm{g}$, based on a single observation, and the same quantity using the combining set of observations, from Eq.~(\ref{combination}). Figure \ref{f:combined-sigmoid} summarises the results. It shows a monotonic increase in the value of the $95\%$ lower limit calculated from the combined PDF, compared to each single $95\%$ lower limit. It is interesting to note that how much we know about $\lambda_\mathrm{g}$ (here quantified by $\lambda_\mathrm{g}^{95\%}$) never decreases as more information (data) are collected. If a ``bad'' observation is recorded (with an individual limit on $\lambda_\mathrm{g}$ significantly worse than the rest of the data), this leaves the overall $95\%$ limit constant. 

\subsection{Combining observations in practice}
\label{s:comb-obs}

In practice, we are dealing with a finite number of samples from the posterior distributions relevant to each detection, rather than with an analytical function. This introduces certain complications when producing the combined PDF on the $\lambda_\mathrm{g}$ parameter, which we have tackled using the procedure that we describe below. The technique that we propose here is applied to the specific example of the graviton's Compton wavelength but can be applied to any parameter in any theory.

To state the problem, we wish to find an appropriate approximation to the posterior PDF for each {observation} which we can multiply together in order to get the combined PDF, {Eq.~(\ref{combination}), through Eq.~(\ref{e:comb1})}. A typical approach is to categorise the samples into a series of bins, creating an histogram which approximates the underlying PDF. A histogram $\mathbf{m}$ is a set of $k$ integers, $\mathbf{m}=(m_1,\ldots,m_k)$, which register the number of samples falling into $k$ independent categories. As a histogram approximates a probability distribution, which we know is normalised, we can approximate the probability in each bin by dividing each count by $n=\sum_{i=1}^k m_i$.
This process can introduce a large amount of noise to the distributions, as there are likely to be very few posterior samples falling into the bins of low probability density, which occurs near the region of the $95\%$  limit (or any other large probability interval) in which we are interested. A na\"{i}ve approach to combining these histograms would be to multiply the (normalised) results in each bin, but if one of the histograms contains zero samples in a bin, then all combined results produced from this distribution will also register zero in that bin no matter how many samples may appear there in the other histograms used in the combination. This problem arises because of the treatment of the normalised count in each bin as if it were the actual probability in that bin. In fact, given a particular histogram, we can calculate the actual probability distribution for the probability in each bin, using the Dirichlet distribution. This allows us to avoid the probability in any one bin going to zero, although that may still be the most likely value given the histograms.

The probability of obtaining a given histogram of $k$ bins from a set of $n$ items is governed by a multinomial distribution:
\begin{align}
p(\mathbf{m};\mathbf{p}) & = \frac{n!}{m_1!\cdots m_k!}\prod_{i=1}^{k}p_i^{m_i}\,, 
\\
n & = \sum_{i=1}^k m_i\,.
\end{align}
Thus we can alternatively describe an histogram using the probabilities $ \mathbf{p} = (p_1, \ldots, p_k)$ associated with each bin of the histogram itself. 
The probability distribution of the probabilities entering the definition of the multinomial distribution is described by the Dirichlet
distribution. In fact the probability density for the variables $ \mathbf{p} = (p_1, \ldots, p_k)$ 
given the histogram $ \mathbf{m} = (m_1, \ldots, m_k)$ is 
\be
p(\mathbf{p};\mathbf{m}) = \frac{1}{Z(\mathbf{m})} \prod_{i=1}^k p_i^{m_i-1} \,,
\ee
where $ p_1, \ldots, p_k \ge 0$, $\sum_{i=1}^k p_i = 1$ and $ m_1, \ldots, m_i > 0$. In this context, the parameters $m_i$ can be interpreted as ``prior observation counts'' for events governed by $p_i$. Furthermore, in the limit $m_i=0$ the Dirichlet distribution is a conjugate non-informative prior for the multinomial distribution. 
The normalisation constant $Z(\mathbf{m})$ is given by 
\be
Z(\mathbf{m}) = \frac{\prod_{i=1}^k \Gamma(m_i)}{\Gamma( \sum_{i=1}^k m_i )}\,,
\ee
where $\Gamma$ is the gamma function. We use the Dirichlet distribution to calculate the probability $p_1, \ldots, p_k$ associated to each of the $k$ bins given the current histogram $\mathbf{m}$. Thanks to this procedure no bin is ever assigned a zero probability. Therefore, to calculate the combined PDFs, Eq.(\ref{combination}), we multiply ``probability histograms", \emph{i.e.} the Dirichlet distributions, describing the probabilities associated with each bin given each histogram, instead of the multinomial distributions 
corresponding to each histogram. The combined PDF can also be thought as the joint probability distribution of the probability that each bin has to receive a future point, given all the previous observations.

An alternative approach to a very similar problem is presented in Ref. \cite{Mandell:2010}. However, 
the method presented there may be unsuitable to estimate the probability density functions of the parameters 
in regions of the parameters space where they are close to zero. 
Given the typical behavior of $p(\lambda_\mathrm{g}|d)$ (\emph{e.g.} Fig. \ref{f:lambda-pdf}), these regions are 
critical for our goal, see Eq.~(\ref{e:95limit}). We stress again that the technique that
we implement here does not rely on any form of approximation or smoothing of the original, discrete PDF.
%
%
\begin{figure}[!ht]
\resizebox{\columnwidth}{!}{\includegraphics{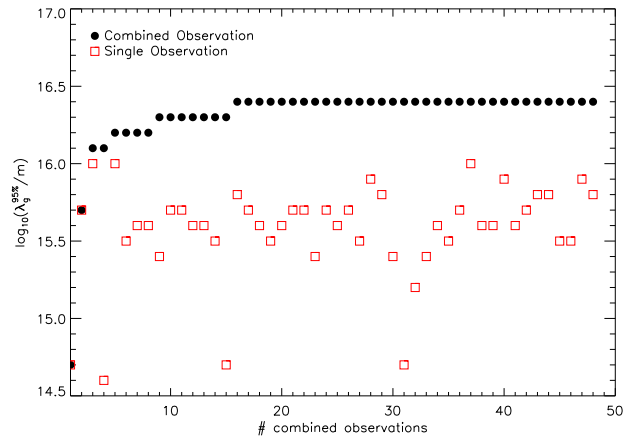}}
\caption{The $95 \%$ lower limit on $\log_{10}(\lambda_\mathrm{g})$ using the detection of 50 inspiral signals.  The plot shows $\lambda^{95 \%}_\mathrm{g}$ for each individual observation (red squares) and for the combined observations (black solid circles). Notice that the combined lower limit does not change much after 10 detections of approximately the same quality (\emph{i.e.} same lower limit on $\lambda_\mathrm{g}$) are combined.}
\label{lambda:combination}
\end{figure}
%
%

We have applied the method described above to a set of $50$ observations of gravitational waves modelled using General Relativity and added, in the same way as reported in the previous Section, to Gaussian and stationary noise following the spectral density of second generation instruments. The parameters of the injections were chosen randomly to create a distribution of signal-to-noise ratios, and we have analysed the data using the MG model. In the analysis the prior range for $p(\lambda_\mathrm{g}|{\cal H}_\mathrm{MG})$, see Eq.~(\ref{lambda:prior}), was $14.5 \le \log_{10}(\lambda_\mathrm{g}/\mathrm{m}) \le 20.5$. For this specific simulation we used a higher number of live points in the nested sampling algorithm than before, specifically $10^4$, in order to generate more accurate PDFs, in particular in regions in which the PDF is close to zero. Figure~\ref{lambda:combination} shows the result of the analysis, where both the $95 \%$ lower limit on $\lambda_\mathrm{g}$ from each individual observation and the combined result, obtained by applying Equation~(\ref{combination}), are reported. The results confirm the behavior that one expects and that we have shown using the proof-of-concept example described in Section~\ref{s:comb-sig} and summarised in Figure~\ref{f:combined-sigmoid}. 
%
%
\begin{figure}[t]
\resizebox{\columnwidth}{!}{\includegraphics{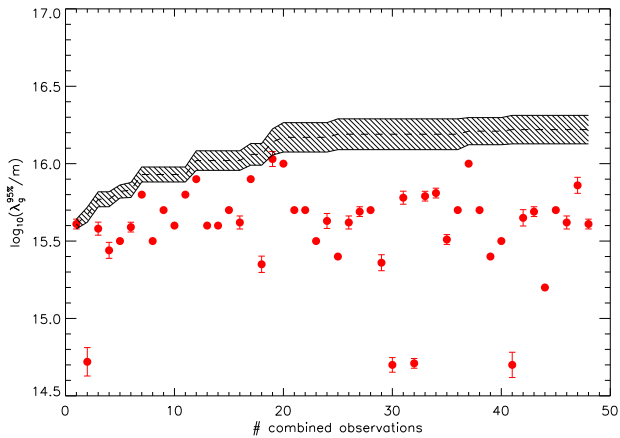}} 
\caption{The shaded region is the $1\sigma$ confidence region over the $95 \%$ lower limit on $\log_{10}(\lambda_\mathrm{g})$ obtained combining $100$ realisations of the single observation PDFs. The order of the observations is held fixed. The circles are the $95 \%$ lower limits on $\log_{10}(\lambda_\mathrm{g})$ for each single observations. The error bars are the $1\sigma$ confidence interval of each point.}
\label{f:combined-samp-perm}
\end{figure}
%
%
Given the discrete nature of the process, the sampling of the original 
PDFs might influence the results presented in Fig.\ref{lambda:combination}. We tested the robustness of our combination method against this effect by studying the scatter introduced on $\lambda_\mathrm{g}^{95 \%}$ by  randomly resampling the nested sampling output chains which yield the marginalised posterior PDFs. For a given single observation, we randomly re-sample the output chains of the nested sampling algorithm to produce 100 marginalised PDFs of $\lambda_\mathrm{g}$. Using the procedure that we have described earlier in the Section, for each of the marginalised PDFs we evaluate $\log_{10}(\lambda_\mathrm{g}^{95\%})$, and then compute the sample mean of the single observation $\log_{10}(\lambda_\mathrm{g}^{95\%})$'s and the spread of these values, in particular we consider the interval over which 68\% of the $\log_{10}(\lambda_\mathrm{g}^{95\%})$'s fall; with slight abuse of terminology, we call this the ``$1\sigma$ interval". We repeat this procedure for each of the individual detections. For a fixed order of the observations, we then draw randomly a value of $\lambda_\mathrm{g}^{95\%}$ coming from the single observation and produce the combined 95\% limit on $\lambda_\mathrm{g}$; we repeat the procedure 100 times to compute a sample mean and a $1\sigma$ interval in the same way as above. The results are summarised in Figure~\ref{f:combined-samp-perm}.

The results of this test confirm that our procedure is robust. We do observe some variations in the combined value of $\log_{10}(\lambda_\mathrm{g}^{95\%})$, but this is restricted to $\approx 0.1$. Moreover, the trend of the mean is consistent with that presented both in Figures~\ref{f:combined-sigmoid} and~\ref{lambda:combination}, and show a clear improvement (well beyond the uncertainty region) with respect to single observations. We are therefore confident in the histogram method of combining multiple independent observations of a constant underlying parameter; it can be applied to any experiment detecting even a small number of gravitational wave events.

Our study already shows a significant improvement on the value of $\lambda_\mathrm{g}^{95\%}$ after the combination of only $5$ PDFs of $\lambda_\mathrm{g}$, which turns out to be larger (by a factor of a few) than the value that any single experiment yields. It is therefore clear that the ability of collectively using the results from many detections shall provide a powerful tool to tighten the observational constraints on the relevant parameters and physical quantities.

\section{Conclusions}\label{s:conclusions}

We have developed a statistical framework and an analysis approach to discriminate among theories of gravity using gravitational waves observations of binary systems. This is a general framework and analysis pipeline that can be straightforwardly extended to any source.  Our analysis strategy is based on Bayesian inference; using a nested sampling algorithm, we can compute the evidence of each theory and then the Bayes factors between them. As a by-product of the analysis, we are also able to compute marginalised posterior probability functions of each model parameter. This approach not only provides a rigorous mathematical quantification of the relative probability of two (or more) theories of gravity given prior information and observations, but also allows the exploration of correlations and bias on the process of parameter estimation in addition to purely statistical errors. We are able to quantify the bias introduced in the analysis if the assumed model does not actually correspond to the "real world" one.

In this framework, we have also introduced the combination of the results from observations of an arbitrary number of sources as a natural extension of the method that enables more powerful inference and therefore better constraints. We consider the fundamental conceptual issues, and the practical ones that derive from dealing with manipulating sampling distributions with few data points.

As a proof-of-principle, we have applied this analysis strategy  to simulated data sets from second generation instruments containing gravitational waves generated during the inspiral of non-spinning compact binary systems in General Relativity and in a ``massive-graviton'' theory of gravity. The focus of this paper has been the methodology, rather than the ability of future experiments to test massive-graviton theories, due to the simplifications that we have made on the gravitational waveform. However, we have derived results that complement past studies that are solely concerned with estimating the expected statistical errors using approximations of the variance-covariance matrix. In fact, the calculation of the evidence suggests that second generation instruments in general will not have sufficient sensitivity to favor a massive-graviton theory over General Relativity, given the existing constraint on the Compton wavelength of the graviton $\lambda_\mathrm{g} \ge 2.8\times 10^{15}$m.  However, we have shown that they may put a limit on $\lambda_\mathrm{g}$ that is a few times more stringent than current bounds. Furthermore, the combination of multiple detections will lead to significant improvements in such a bound.

This study should be regarded as the starting point for more comprehensive investigations, that extend beyond the simple case of non-spinnining inspiral waveforms and tackle in a comprehensive fashion the theories of gravity that are considered viable alternatives to Einstein's theory. It is essential that these aspects (including an end-to-end analysis pipeline to be used by the time advanced instruments are on-line) are fully addressed and we plan to come back to some of them in the future.
  
\section*{Acknowledgements}
We would like to thank F.~Pretorius, N.~Yunes and I.~Mandel for useful discussions. WDP would like to thank B.~Aylott for help in setting up the code. This work has been supported by the UK Science and Technology Facilities Council. The numerical simulations were performed on the Tsunami cluster of the University of Birmingham.

%
%

\end{document}